\PassOptionsToPackage{dvipsnames}{xcolor}
\documentclass[aps, prd, twocolumn, amsmath, floats, floatfix, superscriptaddress, nofootinbib]{revtex4-2}
\usepackage[english]{babel}
\addto\captionsenglish{}

\usepackage{bm}
\usepackage{amsmath}
\usepackage{amssymb}
\usepackage{graphicx}
\usepackage{algorithm}
\usepackage{algpseudocode}

\usepackage{cancel}
\usepackage{booktabs}
\usepackage{multirow}

\usepackage{array}[=2016-10-06]
\usepackage{tabularx}

\usepackage{siunitx}
\usepackage{braket}
\usepackage{makecell}
\usepackage{appendix}
\usepackage{placeins}
\usepackage{ragged2e}

\usepackage[dvipsnames]{xcolor}
\graphicspath{{figures/}}

\usepackage[hidelinks]{hyperref}

\newcolumntype{C}{>{\centering\arraybackslash}X}
\newcolumntype{L}{>{\raggedright\arraybackslash}X}

\algrenewcommand\alglinenumber[1]{}

\algrenewcommand\algorithmicrequire{\textbf{Require:}}
\algrenewcommand\algorithmicensure{\textbf{Ensure:}}
\algrenewcommand\algorithmicforall{\textbf{for all}}
\algrenewcommand\algorithmicif{\textbf{if}}
\algrenewcommand\algorithmicthen{\textbf{then}}
\algrenewcommand\algorithmicelse{\textbf{else}}
\algrenewcommand\algorithmicend{\textbf{end}}
\algrenewcommand\algorithmicfor{\textbf{for}}
\algrenewcommand\algorithmicwhile{\textbf{while}}
\algrenewcommand\algorithmicreturn{\textbf{return}}

\begin{document}

\title{Kerr Quasinormal Modes without Variable Separation: A Two-Dimensional Hyperboloidal Teukolsky Solver with Physics-Informed Neural Networks}


\author{Antonio Ferrer-Sánchez}
\email{Antonio.Ferrer-Sanchez@uv.es}
\affiliation{Intelligent Data Analysis Laboratory (IDAL), Department of Electronic
Engineering, ETSE-UV, University of Valencia, Spain}

\author{Daniela D. Doneva}
\affiliation{Departamento de Astronomía y Astrofísica, Universitat de València, Avda.~Vicent Andrés Estellés 19, Burjassot, 46100, Valencia, Spain}
\affiliation{Theoretical Astrophysics, Eberhard Karls University of T\"ubingen, 72076 T\"ubingen, Germany}

\author{José D. Martín-Guerrero}
\affiliation{Intelligent Data Analysis Laboratory (IDAL), Department of Electronic
Engineering, ETSE-UV, University of Valencia, Spain}
\affiliation{Valencian Graduate School and Research Network of Artificial Intelligence (ValgrAI), Spain}

\author{Stoytcho S. Yazadjiev}
\affiliation{Department of Theoretical Physics, Faculty of Physics, Sofia University, Sofia 1164, Bulgaria}
\affiliation{Institute of Mathematics and Informatics, Bulgarian Academy of Sciences, Acad. G. Bonchev St. 8, Sofia 1113, Bulgaria}

\author{Roberto Ruiz de Austri-Bazan}
\affiliation{Instituto de Física Corpuscular CSIC-UV, c/Catedrático José Beltrán
2, Paterna, 46980, Valencia, Spain}

\author{Yolanda Vives-Gilabert}
\affiliation{Intelligent Data Analysis Laboratory (IDAL), Department of Electronic
Engineering, ETSE-UV, University of Valencia, Spain}

\author{José A. Font}
\affiliation{Departamento de Astronomía y Astrofísica, Universitat de València, Avda.~Vicent Andrés Estellés 19, Burjassot, 46100, Valencia, Spain}
\affiliation{Observatori Astronòmic, Universitat de València, Catedrático José Beltrán
2, Paterna, 46980, Valencia, Spain}




\begin{abstract}
We use physics-informed neural networks (PINNs) to solve the gravitational quasinormal-mode (QNM) eigenvalue problem for Kerr spacetime directly in the two-dimensional hyperboloidal formulation of the Teukolsky equation. This formulation does not require separation of variables and thus retains the coupled radial--angular structure. Such a scheme provides a prototype for calculating the QNMs of beyond-Kerr black holes for which the perturbation equations are non-separable. Sequences with increasing angular momentum are constructed, reaching close to the extremal limit.  We focus on the fundamental modes $(\ell,m,n)=(2,0,0)$, $(2,1,0)$, $(2,2,0)$, $(3,3,0)$ and $(4,4,0)$, together with the first two overtones $(2,2,1)$ and $(2,2,2)$. After training, the frequency at every spin is refined by solving the discretized equation for it, so that it depends much less on where the stochastic training happens to stop. Comparison with continued-fraction values shows that every reported real and imaginary frequency component of the production network remains below $0.23\%$ error, with a median deviation of $2\times10^{-4}\%$. For the fundamental modes, both components agree within about $10^{-3}\%$ up to $a/M=0.97$, and the real parts within $5\times10^{-4}\%$ at every spin. Sub-percent accuracy is maintained in the near-extremal regime, where the damping rate becomes small and the modes are longest-lived; there the damping errors grow, and the largest, $0.23\%$, belongs to the $(2,2,1)$ overtone. The results establish a non-spectral numerical route to multidimensional black-hole perturbation eigenproblems which does not match the substantially higher precision of dedicated Kerr solvers but offers greater flexibility and requires less analytical pre-processing. The proposed approach is designed to be applicable to other backgrounds whose perturbation equations do not separate.
\end{abstract}




    \maketitle 
    
\section{Introduction}
\label{sec:Introduction}

The ringdown of perturbed black holes is one of the cleanest regimes in which strong-field gravity can be tested. After the nonlinear merger stage, the remnant black hole relaxes toward equilibrium through a superposition of damped oscillations, whose complex frequencies are known as quasinormal modes (QNMs). In general relativity (GR), the QNM spectrum of an isolated Kerr black hole is fixed entirely by the mass $M$ and angular momentum $J$ of the remnant. Therefore, measuring these frequencies provides a direct route to black-hole spectroscopy and to consistency tests of the Kerr hypothesis~\cite{Kokkotas_1999,Nollert_1999,Berti_2009}. From a theoretical point of view, QNMs are the eigenvalues of a linear perturbation problem with outgoing behavior at future null infinity and ingoing behavior at the future event horizon.

In the Kerr geometry, linear perturbations obey the Teukolsky master equation, which separates into a radial and an angular ordinary differential equation (ODE) coupled by the complex frequency and by an angular separation constant~\cite{Teukolsky_1972,Teukolsky_1973}. Separability, however, is a special property of the Kerr solution. It relies on the algebraic structure of the background and is not guaranteed in more general modified gravity rotating spacetimes. In such scenarios, the perturbation problem may remain a genuinely multidimensional partial differential equation (PDE) rather than a pair of separable ODEs~\cite{Chung_2024}. This motivates QNM solvers that do not introduce an angular separation constant and that can treat the eigenvalue problem directly in two dimensions.

Early studies in this direction evolved the Teukolsky equation directly as a (2+1)-dimensional PDE after decomposing only the azimuthal dependence~\cite{Krivan_1997,Harms_2013}. More recent developments instead treat the Kerr problem in the frequency domain and solve either the full metric perturbations using spectral methods~\cite{Chung_METRICS_2024,Blazquez_Kerr_2024}, or the Teukolsky equation in a hyperboloidal formulation~\cite{Assaad_2025}. Moreover, the genuinely nonseparable gravito-electromagnetic perturbations of a Kerr--Newman black hole have already been addressed in Ref.~\cite{Dias_2015}. An alternative, hybrid spectral approach, where the spectral coefficients are obtained through a physics-informed neural network (PINN), was developed in Refs.~\cite{Pombo_2026,Pombo_Coupled_2026}.

One of the most important implications of these developments is in theories beyond GR where an analog of the separable Teukolsky equation often does not exist. Analytical studies of the perturbation equations in modified gravity were performed in a series of papers~\cite{Wagle_2023,Hussain_Zimmerman_2022,Cano_2023}. The first breakthrough in this direction was marked by the calculation of the QNM frequencies for rapidly rotating black holes in scalar--Gauss--Bonnet gravity~\cite{Chung_2024,Chung_Yunes_sGB_2024,Blazquez_EGBd_2025}. The spectral approach, though, possesses subtle points~\cite{Chung_2024}. This motivates the search for alternative methodologies which are more straightforward to apply and less sensitive to initial assumptions and regularization procedures.

PINNs offer a different discretization of the same continuum problem, in which the unknown field is represented by a trainable network instead of an expansion in a prescribed basis~\cite{Raissi_2019}. The potential advantage is that neither a separation ansatz, nor a global basis, nor a differentiation matrix is required, and the eigenvalue can be optimized together with the field. In numerical-analysis terms, the network is a differentiable trial function for the unknown field. The method minimizes the Teukolsky residual at a set of points in the compactified domain, and automatic differentiation supplies the required coordinate derivatives~\cite{Raissi_2019}. The complex frequency is an additional unknown of the same nonlinear optimization problem. Thus, the PINN changes the representation and residual-evaluation procedure, but not the continuum Teukolsky equation or its QNM regularity conditions. Machine learning techniques, in PINNs in particular, have already been used in black-hole QNM calculations~\cite{Ovgun_2021,Cornell_2022,Luna_2023,Luna_2024,Patel_2024}. In particular, Ref.~\cite{Luna_2023} solved the separated Kerr Teukolsky equations with two networks for the radial and angular functions while treating both the complex frequency $\omega$ and the angular separation constant $A_{\ell m}$, both defined in Sec.~\ref{subsec:theory}, as trainable parameters. PINNs have also been applied to QNMs in modified-gravity backgrounds, where perturbation equations and backgrounds can be substantially more involved~\cite{Luna_2024}. The present work addresses a different and more challenging numerical problem. We solve the gravitational Kerr Teukolsky QNM problem with spin-weight $s=-2$ using a non-separated hyperboloidal PINN; this paves the way, in turn, to the solution of more complicated problems in the field of modified gravity theories.

The hybrid spectral--PINN calculations discussed above~\cite{Pombo_2026,Pombo_Coupled_2026} retain global polynomial expansions for the unknown fields. Here, by contrast, the regular hyperboloidal field is represented directly in coordinate space, and neither the field representation nor the residual evaluation uses a spectral basis or differentiation matrix. The aim of our approach is, thus, not to compete with the precision of dedicated spectral eigensolvers, but to establish a compact non-spectral baseline for the unseparated hyperboloidal Teukolsky problem. Such a representation does not require the domain to be adapted to a tensor-product basis, nor the operator coefficients to be smooth enough for a global expansion to converge rapidly: collocation points can be placed anywhere in the compactified square, and every derivative entering the residual is obtained by automatic differentiation rather than a differentiation matrix. This is an advantage precisely in the settings that motivate a non-separated solver, such as backgrounds or perturbation coefficients that are known only numerically.

Hyperboloidal compactification provides a particularly suitable starting point for such a formulation. The authors in~\cite{Assaad_2025} recently used this idea, together with an azimuthal $m$-mode decomposition, to formulate Kerr QNMs as a two-dimensional hyperboloidal eigenvalue problem. In~\cite{Assaad_2025} the resulting operator was solved with Chebyshev spectral discretization. Here we use the same continuum formulation, whose radial-fixing coefficients are collected in Appendix~\ref{app:hyperboloidal_operator}, but evaluate the differential residual by automatic differentiation on a neural representation of the regular field. Before adopting this framework, our own first attempts at the two-dimensional problem used compactified Boyer--Lindquist coordinates, $x=r_+/r$ and $u=\cos\theta$, and imposed the ingoing and outgoing behavior through prescribed analytic radial and angular prefactors; those preliminary calculations already yielded viable QNM frequencies. We present the hyperboloidal formulation instead because it enforces the same conditions with no Leaver-type asymptotic prefactor, leaving the represented field regular at both compactified radial boundaries. In addition, this independence from a closed-form asymptotic behavior is what allows the same construction to be carried over to backgrounds for which such analytic factors are not known; the slicing and the regularizing rescaling still carry the asymptotic phases, through the surface gravities and angular velocities of both horizons (Appendix~\ref{app:hyperboloidal_operator}), but they need no closed-form asymptotic solution beyond them. That earlier formulation and its comparison with the hyperboloidal one are summarized in Appendix~\ref{app:prefactor_formulation}.

The current study encompasses seven branches, namely
%
\begin{equation}
    \begin{aligned}
        (\ell,m,n)\in\{&(2,0,0),(2,1,0),(2,2,0),(3,3,0),\\
        &(4,4,0),(2,2,1),(2,2,2)\},
    \end{aligned}
    \label{eq:intro_mode_set}
\end{equation}
all evaluated on the same spin grid from $a/M=0$ to $a/M=0.9998$, where $a/M=J/M^{2}$ is the dimensionless Kerr spin. Throughout, a branch means one such label $(\ell,m,n)$ followed continuously in spin, that is, a single curve $\omega(a)$ that starts from the corresponding Schwarzschild mode at $a/M=0$ and is tracked from one grid point to the next up to the near-extremal end of the grid. The numerical method combines residual minimization with spin continuation and branch tracking. After the Schwarzschild initialization, branch propagation uses the computed residuals, accepted frequency history, field-continuity diagnostics, and predictors built on the analytic zero-damping asymptote of Kerr, but no reference frequencies. We report results spanning the above seven $(\ell,m,n)$ branches, examining their near-extremal damping, quality factors, and waveform implications. 

At the near-extremal end of the grid the residual is almost flat along the damping direction, and a gradient-based optimizer stops at frequencies that differ between runs and processors; the same happens at some moderate spins. Two parts of the method resolve this.  First, at every spin the frequency found by training is refined by Gauss--Newton steps on the residual of a field that has been fitted further at fixed frequency, with the response of the output layer projected out in the manner of variable projection~\cite{Golub_Pereyra_1973,Golub_Pereyra_2003}. Second, the branches with $m\ge2$ are solved near extremality in the Cauchy-horizon-fixing gauge of Ref.~\cite{Assaad_2025}, in which the near-horizon field is smoother.

The organization of this paper is as follows: Section~\ref{subsec:theory} states the Teukolsky equation and its hyperboloidal formulation. Section~\ref{sec:Methodology} describes the numerical discretization, the continuation strategy, and the branch diagnostics. Our results are discussed in  Sec.~\ref{sec:Results}, which reports the QNM frequency spectra for all seven branches together with a comparison with continued-fraction reference values. We close the paper in Sec.~\ref{sec:Conclusions}, where we summarize our conclusions and outline proposals for further research. Complementary information is provided in three appendices. Throughout the paper we employ geometrized units $c=G=1$.

\section{Kerr perturbations and the Teukolsky equation}
\label{subsec:theory}

Linear perturbations of the Kerr spacetime can be described in terms of a
single complex master function $\Psi_s$ of spin weight $s$ obeying the
Teukolsky equation~\cite{Teukolsky_1972,Teukolsky_1973}. In
Boyer--Lindquist coordinates $(t,r,\theta,\phi)$, its explicit vacuum form is
\begin{align}
    0={}&
    \left[\frac{(r^2+a^2)^2}{\Delta}-a^2\sin^2\theta\right]
    \partial_t^2\Psi_s
    +\frac{4Mar}{\Delta}\partial_t\partial_\phi\Psi_s
    \nonumber\\
    &+\left[\frac{a^2}{\Delta}-\frac{1}{\sin^2\theta}\right]
    \partial_\phi^2\Psi_s
    -\Delta^{-s}\partial_r\!\left(\Delta^{s+1}
    \partial_r\Psi_s\right)
    \nonumber\\
    &-\frac{1}{\sin\theta}\partial_\theta\!\left(\sin\theta\,\partial_\theta\Psi_s\right)
    +\left(s^2\cot^2\theta-s\right)\Psi_s
    \nonumber\\
    &{-2s\left[\frac{M(r^2-a^2)}{\Delta}
    -r-ia\cos\theta\right]\partial_t\Psi_s}
    \nonumber\\
    &{-2s\left[\frac{a(r-M)}{\Delta}
    +\frac{i\cos\theta}{\sin^2\theta}\right]\partial_\phi\Psi_s,}
    \label{eq:intro_teukolsky_operator}
\end{align}
where $\Psi_s$ can be related to the metric perturbation and
\begin{equation*}
    \begin{aligned}
    &\Delta=r^2-2Mr+a^2=(r-r_+)(r-r_-),\\
    &r_\pm=M\pm\sqrt{M^2-a^2}.
    \end{aligned}
\end{equation*}

The second-order operator in Eq.~\eqref{eq:intro_teukolsky_operator}
involves the background only through the black hole mass $M$ and specific angular momentum  $a=J/M$. The roots of $\Delta$, namely $r_\pm$,
locate the event and Cauchy horizons, respectively.

The conventional frequency-domain treatment exploits the separability
of Eq.~\eqref{eq:intro_teukolsky_operator}. For a mode with complex frequency
$\omega$ and azimuthal number $m$, the standard separated ansatz is
\begin{equation}
    \Psi_s(t,r,\theta,\phi)
    =e^{-i\omega t}e^{im\phi}
    R_{s\ell m}(r)S_{s\ell m}(\theta;a\omega).
    \label{eq:teukolsky_separated_ansatz}
\end{equation}
After decomposing the field into radial and angular factors, the original PDE reduces to a radial ODE for $R_{s\ell m}(r)$ and an angular spin-weighted spheroidal equation for $S_{s\ell m}$. These two sectors are coupled by the complex frequency and by the angular separation constant $A_{\ell m}$, whose Schwarzschild limit is
\begin{equation}
    \lim_{a\to0} A_{\ell m} = \ell(\ell+1)-s(s+1)\,.
    \label{eq:A_schwarzschild_intro}
\end{equation}
The limit in Eq.~\eqref{eq:A_schwarzschild_intro} fixes the angular-eigenvalue normalization used when comparing with the non-rotating limit.

The QNM spectrum is selected by imposing the physical radiative
conditions at the event horizon and at infinity. Namely, the perturbation function $\Psi_s$ should have the form of a purely outgoing wave at infinity and a purely ingoing wave at the horizon. These conditions make the
problem non-Hermitian and select a discrete set of complex frequencies
$\omega_{\ell mn}$, with $\operatorname{Im}(\omega)<0$ for decaying modes
under the convention used here.

This separated formulation underlies high-accuracy methods such as Leaver's
continued fractions~\cite{Leaver_1985},
Wentzel–Kramers–Brillouin (WKB) approximations~\cite{Seidel_Iyer_1990},
asymptotic iteration~\cite{Cho_2012}, and pseudospectral
discretizations~\cite{Jansen_2017}; time-domain evolutions provide independent checks~\cite{Harms_2013}. The continued-fraction method provides the reference spectra
against which our results are benchmarked.

\subsection{Hyperboloidal formulation}

Slices of constant $t$ are poorly adapted to this problem: they accumulate
at spatial infinity and at the bifurcation sphere, and the QNM eigenfunctions
consequently diverge at both ends of the radial domain. Following
Ref.~\cite{Assaad_2025}, we work instead on a hyperboloidal foliation in the
radial-fixing minimal gauge, defined by
\begin{equation}
    t=r_+\left[\tau-H(\sigma)\right],\;\;
    \sigma=\frac{r_+}{r},\;\;
    u=\cos\theta,\;\;
    \phi=\bar\varphi-\bar\chi(\sigma).
    \label{eq:intro_hyperboloidal_map}
\end{equation}

The radial compactification maps the whole black-hole exterior onto
$\sigma\in[0,1]$, with $\sigma=0$ at future null infinity $\mathcal{I}^+$
and $\sigma=1$ at the future event horizon $\mathcal{H}^+$, while the height
function $H$ and the azimuthal function $\bar\chi$ tilt the slices so that
they terminate on those two null surfaces instead of on spatial infinity.
Both functions are collected in Appendix~\ref{app:hyperboloidal_operator}. Near extremality the branches with $m\ge2$ are solved in a second member of the same family, the Cauchy-horizon-fixing gauge (Sec.~\ref{subsubsec:cauchy_gauge}).

The power of this formulation is twofold: first, the two physical QNM boundaries are included directly in the computational domain; second, while the asymptotic conditions at infinity and at the horizon  must be imposed explicitly in a conventional Boyer--Lindquist treatment, in the hyperboloidal formulation they can instead be encoded to be automatically satisfied, as we show below.

The master function can be rescaled into a variable that stays finite on
the compactified domain,
\begin{equation}
    \Psi_s=\Omega\,\Delta^{-s}\,
    (1+u)^{\delta_-/2}(1-u)^{\delta_+/2}\,\bar\Psi_s,
    \qquad \Omega=\frac{\sigma}{r_+},
    \label{eq:intro_regular_field}
\end{equation}
\noindent with $\delta_\mp=|m\mp s|$. The functions $\Omega$ and
$\Delta^{-s}$ carry the power-law behavior of $\Psi_s$ at future null infinity and at the future event horizon (the latter in the Kinnersley tetrad), and the two remaining factors remove the coordinate singularities
on the symmetry axis $u=\pm1$~\cite{Assaad_2025}. The oscillatory behavior at both ends is removed by the height function $H$ and, for $m\ne0$, by the azimuthal function $\bar\chi$. The physical consequence
is that the QNM conditions are no longer boundary conditions to be imposed:
an eigenfunction is singled out by requiring $\bar\Psi_s$ to be regular on
$(\sigma, u)\in[0,1]\times[-1,1]$, and the outgoing and ingoing behavior then follows from
the causal structure of the slicing. It is this regular field, rather than
$\Psi_s$ itself, that defines the hyperboloidal eigenvalue problem
solved below.

Once the time and azimuthal dependence are removed,
\begin{equation}
    \bar\Psi_s=e^{-i\bar\omega\tau}e^{im\bar\varphi}h(\sigma,u),
    \qquad\bar\omega=r_+\omega,
    \label{eq:intro_nonseparated_ansatz}
\end{equation}
\noindent {the problem reduces to determining the regular
two-dimensional field} $h(\sigma,u)$ directly on $\sigma\in[0,1]$,
$u\in[-1,1]$. The ansatz in Eq.~\eqref{eq:intro_nonseparated_ansatz} defines the
only mode decomposition used in the two-dimensional problem. 
No ansatz of the form
$h(\sigma,u)=f(\sigma)g(u)$ is imposed, and no angular separation constant
is optimized. The label $\ell$ identifies each branch by its Schwarzschild
limit and does not introduce a separated angular eigenproblem at nonzero
spin. In both hyperboloidal gauges used here, the regular field satisfies a quadratic eigenvalue problem
in the frequency parameter. 

The explicit form of the Teukolsky equation, obtained following the transformations presented in this section, is given explicitly in Appendix~\ref{app:hyperboloidal_operator}. This is the form that will be used for the numerical solution of the problem through PINN. The numerical realization is presented in Sec.~\ref{subsec:PI_res}.

\section{Methodology}
\label{sec:Methodology}
Following the hyperboloidal formulation, the transformed Teukolsky equation has to be solved for the regular two-dimensional perturbation
\begin{equation}
    h=h(\sigma,u),\qquad(\sigma,u)\in[0,1]\times[-1,1],
\end{equation}
and the complex QNM frequency $\omega$. 
Spins are reported as the dimensionless quantity $a/M=J/M^{2}$, bounded by the Kerr limit $a/M<1$, and frequencies as the corresponding dimensionless quantity $M\omega$, following the convention of the reference tables; the hyperboloidal operator itself is written with the internal normalization $M=1/2$. Furthermore, we focus on gravitational spin weight $s=-2$. Based on gravitational-wave observations and numerical relativity simulations, we chose to calculate the set of QNMs listed in Eq.~\eqref{eq:intro_mode_set}.

\begin{figure*}[!t]
    \centering
    \includegraphics[width=2.0\columnwidth]{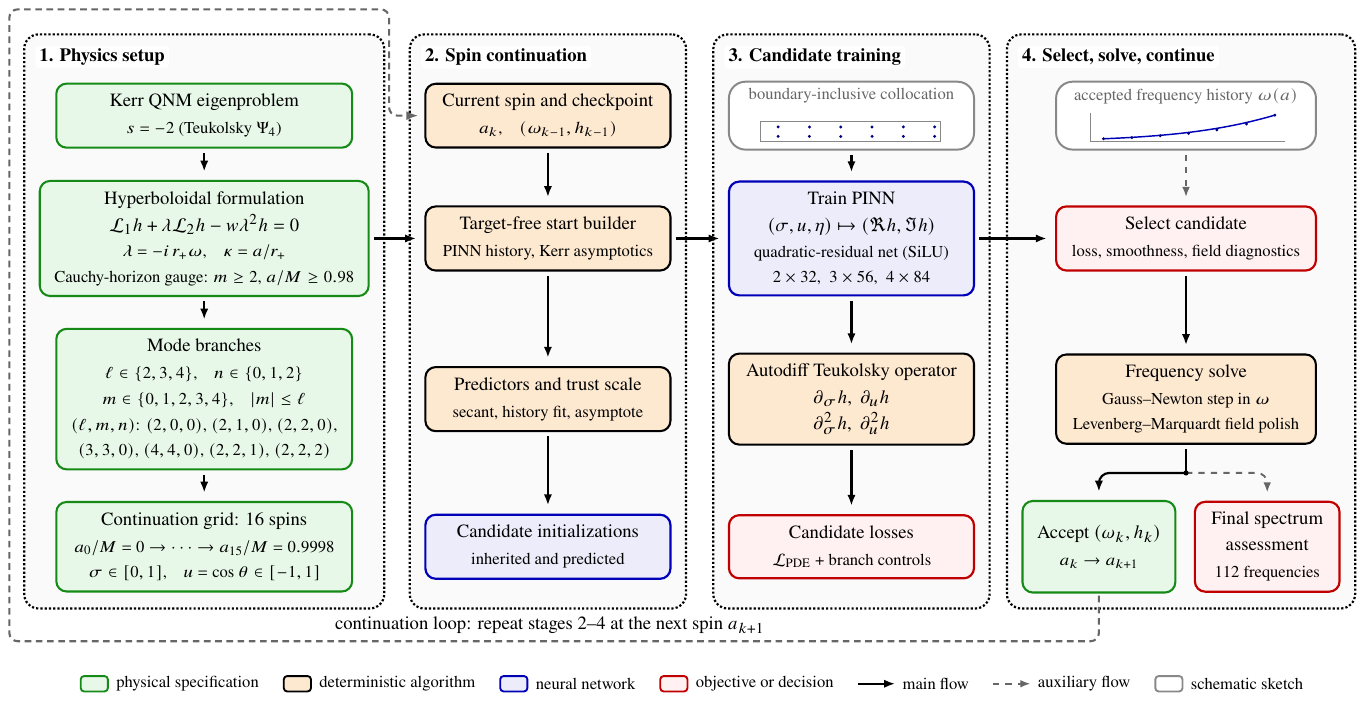}
    \caption{Hyperboloidal spin-continuation pipeline. At each spin a frequency solve, started from the frequency of the selected candidate (or an extrapolated one for $(2,1,0)$ at $a/M\ge0.996$, Algorithm~\ref{alg:pipeline}), sets the frequency, and the accepted eigenpair starts the next spin. The network receives the horizon-adapted coordinate $\eta$ of Eq.~\eqref{eq:horizon_coordinate} as a third input. Apart from the starting values at $a/M=0$, reference frequencies enter only the final assessment.}
    \label{fig:methodology_pipeline}
\end{figure*}

Figure~\ref{fig:methodology_pipeline} summarizes the methodology of the calculation as a pipeline of four stages, arranged so that the physical eigenproblem, its numerical solution, and its independent assessment remain separate, and Algorithm~\ref{alg:pipeline} lists the steps S1--S6 carried out at every spin. The first stage, \textit{physics setup}, collects everything that is fixed before any training: the spin weight, the hyperboloidal operator, the mode labels of Eq.~\eqref{eq:intro_mode_set}, and the spin grid. Each branch is then built one spin at a time, and the three remaining stages are the loop repeated at every grid point. \textit{Spin continuation} evaluates the operator coefficients at the current spin and takes the eigenpair accepted at the previous spin as the starting trial field and frequency. \textit{Candidate training} optimizes the network parameters together with the two frequency components by minimizing the residual of the transformed Teukolsky equation at collocation points in $(\sigma,u)$; these two stages form step S1. \textit{Select, solve, continue} (steps S2--S6) uses convergence and field-continuity diagnostics to select one candidate, refines its frequency and field, and passes the resulting eigenpair on as the starting point at the next spin. For $(2,1,0)$ at $a/M\ge0.996$, S1 and S2 are skipped and the refinement starts from an extrapolated frequency. No reference frequency enters this loop (the Schwarzschild or eikonal starting values at $a/M=0$ are discussed in Secs.~\ref{subsubsec:initialization} and~\ref{subsubsec:phys_eqs}): only once the continuation over the 16 spin values is complete is the assembled branch compared with the continued-fraction reference spectrum, as reported in Sec.~\ref{sec:Results}.

\begin{algorithm}[H]
\caption{Computation of one branch $(\ell,m,n)$ over the spin grid $a_0<a_1<\dots<a_{15}$ of Eq.~\eqref{eq:a_continuation}.}
\label{alg:pipeline}
\small
\begin{algorithmic}
\Require operator of Eq.~\eqref{eq:pinn_residual}; starting frequencies at $a_0=0$ (Sec.~\ref{subsubsec:initialization}); $N_k$ candidates at $a_k$; $\bm\omega=(\omega_R,\omega_I)$. At $k=0$, S1 and S4 start from a fresh network and S3 from the field of the selected candidate; for $(2,1,0)$ at $a_k/M\ge0.996$, S1 and S2 are skipped and $\bm\omega$ is extrapolated; for $m\ge2$ and $a_k/M\ge0.98$, all steps use the Cauchy-horizon-fixing gauge (Sec.~\ref{subsubsec:cauchy_gauge})
\For{$k=0,1,\dots,15$}
\State \textbf{S1} Train $N_k$ candidates on the loss $\mathcal{L}$ of Eq.~\eqref{eq:total_loss}, from the field accepted at $a_{k-1}$ and from frequencies extrapolated from the accepted ones, held fixed at first (Secs.~\ref{subsec:PI_res} and~\ref{subsec:spin_cont}); each trained candidate also receives S5 (where it applies) and S6
\State \textbf{S2} Among the eligible candidates, select the one with the smallest score of Eq.~\eqref{eq:continuation_score_overview} (Sec.~\ref{subsec:spin_cont}); $\bm\omega\gets$ its frequency
\State \textbf{S3} Frequency solve (Sec.~\ref{subsubsec:frequency_solve}), from the field accepted at $a_{k-1}$:
\Repeat
\State \textit{inner}: polish the weights at fixed $\bm\omega$, Eq.~\eqref{eq:lm_polish}
\State \textit{outer}: $\bm\omega\gets\bm\omega+\Delta\bm\omega$, Eq.~\eqref{eq:varpro_step}
\Until{both relative changes of $\bm\omega$ are below $10^{-4}$}
\State \textbf{S4} Train the network from the field accepted at $a_{k-1}$, with $\bm\omega$ fixed
\If{$n\ge1$ and $0.2\le a_k/M\le0.96$}
\State \textbf{S5} Refine the weights and $\bm\omega$ together (Sec.~\ref{subsubsec:frequency_solve})
\EndIf
\State \textbf{S6} Polish the field at fixed $\bm\omega$; the result is the eigenpair accepted at $a_k$
\EndFor
\State Compare the branch with the reference spectrum (Sec.~\ref{sec:Results})
\end{algorithmic}
\end{algorithm}
In words, S1 produces trial eigenpairs by training (each candidate is already polished, and for the overtones refined as in S5, before it is ranked), S2 keeps the one that best continues the branch, and S3--S6 refine it. The selected candidate supplies the starting frequency of S3; S3 and S4 restart from the field accepted at the previous spin, so that the field carried along the branch is always one that has passed all steps. Here and below, to \textit{polish} a field means to refine the network weights at a fixed frequency by a few deterministic Levenberg--Marquardt steps, Eq.~\eqref{eq:lm_polish}; this lowers the residual well below the level that the stochastic training reaches, without moving the field to another branch. Step S3 then sets the frequency by requiring that the residual of the polished field be stationary with respect to it (Sec.~\ref{subsubsec:frequency_solve}). Step S4 retrains the field at that frequency, S5 lets the overtone frequencies relax once more together with the field, and S6 polishes the field a last time. The acceptance tests act inside these steps: only candidates whose residual is close to the best one are eligible in S2, and S5 and S6 keep their result only if it also lowers the objective on an independent set of points. The pair returned by S6 therefore needs no further test, and it starts the next spin.

\subsection{Numerical representation and normalization}
\label{subsec:NN_rep}

The regular field is represented by a fully connected neural network (NN) as follows. For the present purpose, the NN can be viewed simply as a nonlinear parametrized trial function for the unknown eigenfunction $h(\sigma,u)$, analogous in spirit to choosing a finite-dimensional basis in a conventional numerical discretization. The model receives the coordinates $(\sigma,u)$, together with a horizon-adapted radial coordinate $\eta(\sigma)$ defined below, as input and returns the real and imaginary parts of the complex field, forming
\begin{equation}
    \mathcal{N}_{\Theta}(\sigma,u)
    =
    \mathcal{N}_{R,\Theta}(\sigma,u)
    +i\:\mathcal{N}_{I,\Theta}(\sigma,u),
    \label{eq:nn_ansatz}
\end{equation}
where $\Theta$ denotes the adjustable coefficients (weights) of the neural trial function. These coefficients play a role similar to expansion coefficients in a spectral representation: they are varied until the trial field satisfies the Teukolsky equation to the desired accuracy. Importantly, no numerical QNM eigenfunctions are supplied as training data. Instead, the parameters are determined from the differential equation itself.

The extra input addresses the near-extremal regime. As $a\to M$ the Cauchy horizon $r_-$ approaches the event horizon, and the field varies across a layer of width $\delta=(r_+-r_-)/r_+=1-\kappa^2$, with $\kappa=a/r_+$, next to $\sigma=1$. A network that sees only $\sigma$ represents this layer poorly, because it favors functions that vary smoothly on the scale of the whole interval. The coordinate
\begin{equation}
    \eta(\sigma)=1-\frac{\ln\left[1+(1-\sigma)/\delta\right]}{\ln\left(1+1/\delta\right)}
    \label{eq:horizon_coordinate}
\end{equation}
maps $[0,1]$ onto itself and stretches the layer $1-\delta\le\sigma\le1$ over a fraction $\ln2/\ln(1+1/\delta)$ of the $\eta$ interval. At $a/M=0.9998$, where $\delta\simeq0.04$, the layer takes up $4\%$ of the $\sigma$ interval but $21\%$ of the $\eta$ interval. 

The representation in Eq.~\eqref{eq:nn_ansatz} is our central numerical approach: the field is not expanded in a prescribed global polynomial basis, and the derivatives entering the residual are not obtained from finite-difference stencils or spectral collocation on a fixed grid. Instead, derivatives of the trial function with respect to $\sigma$ and $u$ are evaluated directly by automatic differentiation, as described in Sec.~\ref{subsec:PI_res}.

The QNM frequency is optimized simultaneously and is represented by two additional scalar unknowns corresponding to its real and imaginary parts, $\omega_R$ and $\omega_I\ge0$, with
\begin{equation}
    \omega=\Re(\omega)-i\:|\Im(\omega)|\,\equiv\omega_R-i\,\omega_I.
    \label{eq:omega_param}
\end{equation}
Thus, $\omega$ is treated in precisely the same eigenvalue sense as in a conventional nonlinear eigensolver: both the field and the eigenfrequency are varied until the differential equation is satisfied. The absolute value in Eq.~\eqref{eq:omega_param} restricts the search to $\operatorname{Im}(\omega)\leq0$, corresponding to decaying QNMs for the time dependence adopted in Eq.~\eqref{eq:intro_nonseparated_ansatz}, that is the case for all Kerr QNMs.

Due to the Teukolsky eigenvalue problem being homogeneous, the overall complex scale of the field is arbitrary. To remove this scaling degeneracy and avoid convergence to $h\equiv0$ (trivial solution), a hard single-point normalization condition at an interior anchor $(\sigma_0,u_0)$ is imposed. For convenience,  the normalized model output is chosen to be
\begin{equation}
    \begin{aligned}
    h_{\Theta}(\sigma,u)
    &=
    1+\mathcal{N}_{\Theta}(\sigma,u)
    -\mathcal{N}_{\Theta}(\sigma_0,u_0),\\
    h_{\Theta}(\sigma_0,u_0)&=1,
    \end{aligned}
    \label{eq:hard_anchor}
\end{equation}
\noindent with $(\sigma_0,u_0)=(0.5,0)$. Here $h_{\Theta}$ denotes the field produced by the NN with parameters $\Theta$; henceforth, $h$ denotes this model output. The constraint in Eq.~\eqref{eq:hard_anchor} only fixes the global amplitude and phase. It does not constrain the radial or angular shape of the eigenfunction, but instead merely chooses one representative from the family $c\,h(\sigma,u)$, with arbitrary complex constant $c$, that solves the same homogeneous equation.

\begin{figure}[!t]
    \centering
    \includegraphics[width=\columnwidth]{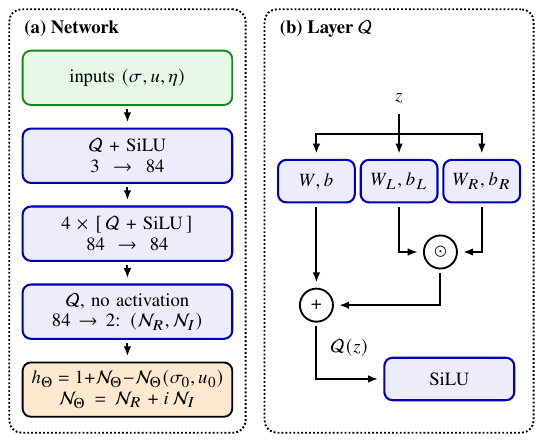}
    \caption{\justifying Quadratic-residual network of the production calculations. (a) The inputs pass through an input layer and four hidden layers of width 84, each a quadratic-residual map $\mathcal{Q}$ followed by a sigmoid linear unit (SiLU) activation, and an output layer without activation; the normalization of Eq.~\eqref{eq:hard_anchor} turns the two outputs into the field $h_\Theta$. (b) One layer, Eq.~\eqref{eq:qres_layer}: a linear map plus the component-wise product $\odot$ of two further linear maps of the same input $z$. The $2\times32$ and $3\times56$ networks differ only in depth and width.}
    \label{fig:architecture}
\end{figure}

The production calculations use a fully connected quadratic-residual network with three inputs and two real outputs~\cite{qres_layers}. The term ``quadratic-residual'' refers only to the chosen parametrization of the trial function: in addition to ordinary linear transformations, the network layers contain quadratic combinations of their internal variables, which provide a more flexible representation of the two-dimensional field. It should not be confused with the quadratic dependence of the Teukolsky eigenvalue problem on $\omega$. The network architecture is a numerical resolution choice, not part of the physical formulation.  From a more machine-learning (ML)-oriented point of view, it might be seen as a set of choices that can be considered beforehand, including aspects such as the number of layers or the number of neurons per layer, as well as the activation functions. Within the field, these choices are commonly known as ``hyperparameters''. There is no single network architecture that is optimal for every problem. However, through our study, we aim to keep the NN as simple as possible while giving it enough learning capabilities to tackle our physical problem, even though it is also important to check the sensitivity of the PINN performance to those hyperparameters  (cf.~Sec.~\ref{subsec:numerical_robustness}). Figure~\ref{fig:architecture} sketches the production network, which has an input layer and four hidden layers of width 84 ($4\times84$). In particular, layer structure, parameter count, initialization, and the two smaller networks used for the robustness study are reported in Appendix~\ref{app:continuation_details} and Table~\ref{tab:architecture_definition}.

\subsection{Physics-informed residual, collocation points and initialization}
\label{subsec:PI_res}

All derivatives entering the hyperboloidal Teukolsky residual are computed by automatic differentiation with respect to the input coordinates. The residual evaluation and the optimization of this subsection constitute stage~3 (\textit{candidate training}) of Fig.~\ref{fig:methodology_pipeline}, that is, step S1 of Algorithm~\ref{alg:pipeline}. In this procedure, derivatives of the trial function are obtained by repeated application of the chain rule through the network representation, rather than by finite differences or differentiation matrices. Given the neural field $h$, the unweighted PDE residual is
\begin{equation}
    \mathcal{R}_{\Theta}(\sigma,u)=\mathcal{L}_{1}[h]+\lambda\mathcal{L}_{2}[h]-w\lambda^{2}h,
    \label{eq:pinn_residual}
\end{equation}
with $\lambda=-ir_+\omega$.
The spin does not appear explicitly in Eq.~\eqref{eq:pinn_residual}; it enters through the coefficient functions of $\mathcal{L}_{1}$, $\mathcal{L}_{2}$ and $w$, which depend on it through the dimensionless combination $\kappa=a/r_+$ used in Appendix~\ref{app:hyperboloidal_operator}. The quadratic dependence on $\lambda$ descends from the second time derivative of Eq.~\eqref{eq:intro_teukolsky_operator}, so the frequency enters as an eigenvalue. The operators $\mathcal{L}_{1}$ and $\mathcal{L}_{2}$ contain the radial, angular, and mixed spin-coupled coefficient functions of the radial-fixing hyperboloidal formulation (see Appendix~\ref{app:hyperboloidal_operator} for further details, which also gives those of the Cauchy-horizon-fixing gauge used near extremality).  The residual is assembled point by point. Each of the operators $\mathcal{L}_{1}$ and $\mathcal{L}_{2}$ acts on $h$ as a linear combination of the field and its first and second derivatives with respect to $\sigma$ and $u$, with coefficient functions that are known in closed form and depend only on the coordinates and on the spin. At a given collocation point, the network returns $h$, automatic differentiation returns those derivatives, the coefficient functions and $w$ are evaluated at the same point, and $\lambda$ is taken from the current values of the two trainable frequency components. Combining these quantities with complex arithmetic yields one complex number per point, and no linear system is formed during training. The network parameters and the frequency are obtained by minimizing a loss function, in the standard PINN sense: a single scalar assembled from averages of residuals over the collocation points.

\subsubsection{PDE residual loss term}

The principal term in the loss function is the mean-squared PDE residual,
\begin{equation}
    \mathcal{L}_{\mathrm{PDE}}=\left\langle\left|W_{\partial}(\sigma,u)\:\mathcal{R}_{\Theta}(\sigma,u)\right|^{2}\right\rangle_{\Omega},
    \label{eq:pde_loss}
\end{equation}
\noindent where $\langle\cdot\rangle_{\Omega}$ denotes the empirical average over collocation points. Those are simply the discrete coordinate locations $(\sigma,u)$ at which the differential equation is evaluated during the numerical solution. They therefore play a role analogous to grid or quadrature points in more conventional PDE discretizations, although they need not form a fixed grid.

The loss in Eq.~\eqref{eq:pde_loss} is the physical residual being minimized: if $\mathcal{L}_{\mathrm{PDE}}$ vanishes, the learned field and frequency satisfy the hyperboloidal eigenproblem at the collocation points. The factor $W_{\partial}$ is an optional numerical residual weight, which we consider to be $W_{\partial}=1$ (unweighted residual). This choice is natural since the hyperboloidal formulation already places the relevant QNM boundaries at the future event horizon and future null infinity, while the regular field redefinition handles the angular-axis behavior. Boundary weighting remains available as a numerical option, but it is not part of the continuum formulation and is not used to impose QNM boundary conditions. The outgoing and ingoing conditions are already contained in the regularity of the rescaled field of Sec.~\ref{subsec:theory}, so the residual is the only equation to be enforced and the collocation set may include the horizon and null infinity themselves. On a constant-$t$ slicing the QNM eigenfunctions instead diverge at both ends of the radial domain, so the outgoing and ingoing behavior has to be supplied explicitly, either by factoring out prescribed asymptotic prefactors, as done in Appendix~\ref{app:prefactor_formulation}, or by matching the solution to an asymptotic expansion at a truncation radius.

\subsubsection{Relative residual loss term}

Our loss function also includes the relative residual term
\begin{equation}
    \mathcal{L}_{\mathrm{rel}}=\left\langle\left|\frac{\mathcal{R}_{\Theta}}{\mathcal{S}_{\Theta}+\epsilon_{\mathrm{rel}}}\right|\right\rangle_{E}.
    \label{eq:relative_loss}
\end{equation}
The term $\mathcal{S}_{\Theta}$ in the denominator  measures the local magnitude of the individual operator terms and is held fixed while each parameter update is computed. Consequently, $\mathcal{L}_{\mathrm{rel}}$ identifies local residual imbalance without changing the Teukolsky operator. It is an auxiliary numerical scaling of the same residual, not an additional field equation or boundary condition. The pointwise scale and endpoint weighting are defined in Appendix~\ref{app:continuation_details}. The term is needed because the leading radial coefficient of the operator (Appendix~\ref{app:hyperboloidal_operator}) vanishes at both radial endpoints, and the coefficients vary by orders of magnitude in between, so an absolute residual would be dominated by the bulk and would tolerate large local errors near the horizon, where the near-extremal eigenfunctions are structured. Thus, $\mathcal{L}_{\mathrm{rel}}$ serves mainly to place residual errors arising in regions with very different characteristic scales on a more comparable footing.

\subsubsection{Auxiliary branch controls}

The unseparated formulation does not introduce the angular separation constant $A_{\ell m}$. Consequently, the optimization landscape may contain several nearby residual-minimizing branches. In more conventional eigenvalue language, this means that several distinct approximate eigenpairs can produce comparably small PDE residuals, and a nonlinear solver may move from one spectral branch to another. To help select regular field branches without imposing a separated angular equation, we use weak mode-selection penalties. They are needed because $A_{\ell m}$, which fixes the angular structure in the separated treatment of Sec.~\ref{subsec:theory}, is absent here, so $\ell$ survives only through the Schwarzschild limit. Drifting toward a neighboring angular harmonic becomes more likely for the higher-$\ell$ branches and toward extremality, where the zero-damped branches with the same $m$, such as $(2,2,n)$, approach a common oscillation frequency set by the rotation of the horizon, $m\Omega_{\rm H}$, with $\Omega_{\rm H}$ being the angular velocity of the horizon (Sec.~\ref{subsec:near_extremal_scaling}), and become hard to differentiate.  At a given spin, we may perform several optimizations instead of one, each started from a different initial guess, and each of them returns one trial pair $(h,\omega)$; these are the candidates. Here, we use the word candidate in a purely numerical sense. Several candidates are therefore attempts at the same physical mode, and the selection step of Sec.~\ref{subsec:spin_cont} keeps one of them as the accepted eigenpair at that spin.

The numerical objective therefore contains several weak auxiliary terms designed to preserve the desired branch during the nonlinear solve.

\subsubsection{Mode-selection loss term}

The first contribution to the mode-selection loss $\mathcal{L}_{\mathrm{mode}}$ penalizes unnecessary angular curvature of the regular field, introduced as
\begin{equation}
    \mathcal{L}_{\mathrm{curv}}=\frac{\left\langle\left|(1-u^{2})\partial_{u}^{2}h\right|^{2}\right\rangle_{\Omega}}{\left\langle |h|^{2}\right\rangle_{\Omega}+\epsilon}.
    \label{eq:curvature_loss}
\end{equation}
The factor $(1-u^{2})$ in Eq.~\eqref{eq:curvature_loss} suppresses artificial contributions at the angular endpoints and makes the penalty act mainly on the interior angular profile. Physically, a spheroidal harmonic of given $(\ell,m)$ has a definite and small number of angular nodes, so rapid oscillations in $u$ correspond to no Kerr mode of the set considered. The factor $(1-u^{2})$ vanishes on the symmetry axis, where regularity is already secured by Eq.~\eqref{eq:intro_regular_field}. 

The second contribution to $\mathcal{L}_{\mathrm{mode}}$ promotes coherence of the angular profile across different radial slices. Let $h_{b}(u)$ denote the normalized angular profile extracted at the $b$th probe value of $\sigma$, after phase alignment with respect to a reference slice. We define schematically
\begin{equation}
    \mathcal{L}_{\mathrm{coh}}=\left\langle\left|h_{b}(u)-\bar{h}(u)\right|^{2}\right\rangle_{b,u},
    \label{eq:coherence_loss}
\end{equation}
\noindent where $\bar{h}(u)$ is the mean phase-aligned profile. Each slice carries an arbitrary amplitude and phase, so every $h_{b}$ is rescaled to unit mean power and phase-rotated onto the reference slice; $\bar{h}(u)$ is the average of the aligned profiles, and only the shape in $u$ is compared. The coherence term in Eq.~\eqref{eq:coherence_loss} does not enforce separability in the form $h(\sigma,u)=f(\sigma)g(u)$ and does not introduce the angular Teukolsky equation. It only discourages incoherent two-dimensional configurations whose angular content changes erratically with radius. In the separated treatment the angular profile is independent of the radius, and the same holds for the exact regular field here, since every factor removed in Eqs.~\eqref{eq:intro_hyperboloidal_map}--\eqref{eq:intro_regular_field} depends on $\sigma$ or on $u$ alone; the method does not impose this property, which need not hold for other backgrounds; an abrupt change between radial slices is therefore an artifact of the trial space. The term is most useful for branches $(3,3,0)$ and $(4,4,0)$.

The total mode-selection loss is, thus,
\begin{equation}
    \mathcal{L}_{\mathrm{mode}}=\mathcal{L}_{\mathrm{curv}}+\mathcal{L}_{\mathrm{coh}}.
    \label{eq:mode_loss}
\end{equation}

The various loss terms discussed thus far are not all active from the beginning of a run, and the order in which they are switched on matters. During the first few hundred epochs the objective contains the residual terms of Eqs.~\eqref{eq:pde_loss} and~\eqref{eq:relative_loss} alone, so that the field and the frequency are driven towards a solution of the Teukolsky equation before any selection is attempted. The weight of Eq.~\eqref{eq:mode_loss} is then raised linearly from zero over a warm-up interval of order $10^{3}$ epochs, up to a value several orders of magnitude below that of the residual terms ($5\times10^{-5}$ to $10^{-4}$, against the unit weight of $\mathcal{L}_{\mathrm{PDE}}$). This sequence is deliberate: the penalties are meant to break the tie between configurations that already give a small residual, and switching them on earlier would shape the field before any approximate eigenpair has been found. Moreover, the overtone terms in the loss, introduced below, are activated in the same way, each with its own start epoch and warm-up length, as detailed in Appendix~\ref{app:continuation_details}.  

The mode-selection weight remains weak and is increased only in the difficult high-spin regimes. This is because at high spin the neighboring branches lie so close together in frequency that a small residual alone is no longer able to identify which one has been found. These auxiliary terms do not impose a separated spheroidal eigenfunction associated with a chosen $\ell$.  The curvature term does not vanish on the exact eigenfunction, because a spheroidal profile has nonzero angular curvature; the coherence term does vanish on the exact Kerr eigenfunction, but it is a penalty, not a constraint. They are biases rather than constraints, so their weight is kept far below that of the residual, and the frequency solve of step S3 (Sec.~\ref{subsubsec:frequency_solve}) uses the plain residual without them.

\subsubsection{Overtone loss terms}

For the $n\ne0$ branches, additional terms can be enabled in the loss function to discourage collapse to a smoother fundamental-like profile.

The first one is a same-branch frequency-control term,
\begin{equation}
    \mathcal{L}_{\mathrm{overtone\;br}}
    =
    \left(\frac{\Re(\omega)-\Re(\omega_{\star})}{s_R}\right)^2
    +
    \left(\frac{\Im(\omega)-\Im(\omega_{\star})}{s_I}\right)^2.
    \label{eq:overtone_branch_loss}
\end{equation}
Here, $\omega_{\star}$ is a local frequency estimate constructed from previously accepted PINN frequencies of the same $(\ell,m,n)$ branch, and $(s_R,s_I)$ are its component-wise continuation scales. Equation~\eqref{eq:overtone_branch_loss} limits a large frequency departure while the overtone field adjusts to the new spin. Physically, the eigenvalue of a given branch varies smoothly with the spin, so a large jump between neighboring grid points indicates a change of branch rather than a better solution of the same one. Frequency control alone, however, does not prevent the learned field from becoming radially too smooth.

The second term is designed to handle the overtone radial shape. It evaluates three dimensionless diagnostics computed directly from the current candidate field:
\begin{align}
    E_{\sigma}&=\frac{\left\langle |\partial_{\sigma}h|^2\right\rangle_{\Omega}}
    {\left\langle |h|^2\right\rangle_{\Omega}+\epsilon},\nonumber\\
    E_{\sigma\sigma}&=
    \frac{\left\langle |\sigma(1-\sigma)\partial_{\sigma}^{2}h|^2\right\rangle_{\Omega}}
    {\left\langle |h|^2\right\rangle_{\Omega}+\epsilon},\nonumber\\
    C_{\sigma}&=
    \frac{\left|\langle h\rangle_{\sigma\leq\sigma_m}-\langle h\rangle_{\sigma>\sigma_m}\right|^2}
    {\left\langle |h|^2\right\rangle_{\Omega}+\epsilon}.
    \label{eq:overtone_shape_diagnostics}
\end{align}
These fixed numerical safeguards, not hyperparameters, prevent zero denominators: Eqs.~\eqref{eq:relative_loss}, \eqref{eq:curvature_loss}, and \eqref{eq:overtone_shape_diagnostics} use $10^{-6}$, $10^{-8}$, and $10^{-10}$, respectively.

Here, $\sigma_m=0.5$ divides the compactified radial domain into inner and outer regions. The three quantities characterize complementary aspects of the radial field structure: $E_{\sigma}$ measures the overall radial-gradient energy, $E_{\sigma\sigma}$ measures the compactified radial-curvature energy, and $C_{\sigma}$ measures the contrast between the mean field values in the two regions. Together, they indicate whether the candidate retains nontrivial radial structure rather than relaxing toward a smoother fundamental-like configuration. We combine them in the schematic loss
\begin{equation}
    \begin{aligned}
        \mathcal{L}_{\mathrm{overtone\;sh}}
        ={}&\frac{1}{1+E_{\sigma}/s_{\sigma}}
        +\eta_{\sigma\sigma}\frac{1}{1+E_{\sigma\sigma}/s_{\sigma\sigma}}\\
        &+\eta_{C}\frac{1}{1+C_{\sigma}/s_{C}}.
    \end{aligned}
    \label{eq:overtone_shape_loss}
\end{equation}

We use $\eta_{\sigma\sigma}=\eta_C=0.5$, $s_\sigma=s_{\sigma\sigma}=1$, and $s_C=1/4$; the overtone-control schedule and overall weights are given in Appendix~\ref{app:continuation_details}.

The loss in Eq.~\eqref{eq:overtone_shape_loss} is large when the learned field is radially too smooth or has too little radial contrast, and it decreases as the current candidate develops radial gradient, curvature, and inner--outer contrast. The loss does not prescribe a node count or a known overtone profile. Instead, it relies exclusively on structural information extracted from the current candidate field. In this way, it gently discourages the optimization from drifting toward an excessively smooth, fundamental-like configuration, while leaving the detailed radial structure of the overtone to be determined through the physics-informed objective. The situation is specific to the overtones: $(2,2,1)$ and $(2,2,2)$ share the angular structure and a real frequency close to that of $(2,2,0)$, equal near extremality, but decay several times faster. As they differ mainly in radial structure, residual minimization alone can slide onto the smoother fundamental mode. For the fundamental $n=0$ branches overtone terms are absent. Hence, their weights are zero for the whole run, at every spin of the sequence, since the collapse they guard against can only happen for an overtone.

\subsubsection{Frequency-control loss term}

In sensitive spin-continuation steps, an auxiliary frequency-control term $\mathcal{L}_{\omega}$ is added to the total loss. It limits departure from a local extrapolation of the accepted same-branch frequencies. The field is first adapted to the new operator while the frequency is held fixed; the frequency is then released, and the control term is scaled relative to the residual objective. This procedure is analogous to a local trust region in nonlinear continuation. It guides the optimizer without supplying a target frequency or modifying the Teukolsky equation. Its complete candidate-specific definition is given in Appendix~\ref{app:continuation_details}.

\subsubsection{Total loss function}

Summarizing, the total loss function minimized in the hyperboloidal runs is
\begin{align}
    \mathcal{L}&=\mathcal{L}_{\mathrm{PDE}}+w_{\mathrm{rel}}\mathcal{L}_{\mathrm{rel}}+\lambda_{\mathrm{mode}}(e)\mathcal{L}_{\mathrm{mode}}\nonumber\\
    &\quad+\mathbf{1}_{n>0}\left[\lambda_{\mathrm{br}}(e)\mathcal{L}_{\mathrm{overtone\;br}}+\lambda_{\mathrm{sh}}(e)\mathcal{L}_{\mathrm{overtone\;sh}}\right] \nonumber \\
    &\quad +\lambda_{\omega}(e)\mathcal{L}_{\omega}+\lambda_{\nabla}\mathcal{L}_{\nabla}.
    \label{eq:total_loss}
\end{align}
Here, $e$ counts the epochs of the gradient-based minimization of Eq.~\eqref{eq:total_loss} over the network parameters $\Theta$ and the two frequency components, so that the weights $\lambda(e)$ realize the activation schedule described above, and the candidate superscript on $\mathcal{L}_{\omega}$ is suppressed. The overtone terms vanish for all fundamental branches and are strengthened only in the numerically more difficult overtone regimes. The frequency-control term is independent of the overtone controls and is activated only where local branch guidance is required. For $m\ne0$ at $a/M\ge0.96$, $\mathcal{L}_{\nabla}=\langle|\partial_\sigma\mathcal{R}_{\Theta}|^2+|\partial_u\mathcal{R}_{\Theta}|^2\rangle_{\Omega}$ is the gradient-enhanced residual term of Ref.~\cite{Yu_2022}, with weight $\lambda_{\nabla}=10^{-2}$, because near extremality the plain residual of a smooth field depends only weakly on the frequency; elsewhere $\lambda_{\nabla}=0$. Only $\mathcal{L}_{\mathrm{PDE}}$ enters with unit weight in the total loss, so the Teukolsky equation remains the single physical constraint; every other term is a subordinate numerical aid, weak by construction, except the frequency-control term, which is normalized to the residual but acts only during training in step S1, and restricted to the regimes where branch identification is delicate. A compact summary of the network initialization, loss weights, optimization, and collocation settings is provided in Appendix~\ref{app:continuation_details}.

\subsubsection{Collocation sampling}

The differential residual is evaluated on a tensor-product set of radial and angular points that is regenerated during the optimization. Resampling is essentially free here, because no matrix is attached to the points, and it serves a purpose: the loss is a sample estimate of the residual over the whole domain, so with a single fixed point set the optimizer can lower the residual at those points while leaving it large in between. Resampling the collocation points every epoch reduces this possibility. The sample includes the exact compactified boundaries because the hyperboloidal equation is regular there; no boundary loss is imposed. Interior points are supplemented by samples near the radial and angular endpoints, and the radial density is increased near the horizon toward extremality to resolve the localized field structure: a share of the radial points, $30\%$ at low spin rising to $52\%$ near extremality, is drawn uniformly in $\eta$. The refinement follows the near-extremal physics: for the co-rotating $m=\ell$ branches the damping tends to zero and the eigenfunction steepens near the horizon~\cite{Yang_2013,Yang_2013_extended}; the same rule is applied to every branch. Appendix~\ref{app:continuation_details} gives the sampling rule and grid sizes.

\subsubsection{Initialization and nonlinear solution}
\label{subsubsec:initialization}


The starting point of the computation is the output of stage~1 of Fig.~\ref{fig:methodology_pipeline} (\textit{physics setup}) feeding the subsequent  stage (\textit{spin continuation}). At $a/M=0$, each branch is initialized near its Schwarzschild frequency; a small list of starting frequencies, fixed in advance rather than drawn at random, is used for the $\ell=3$ and $\ell=4$ branches. These starting values follow the eikonal scaling of the Schwarzschild spectrum, in which the real frequency grows in proportion to $\ell+1/2$, so each higher-$\ell$ branch starts near its own frequency. The overtones start from their Schwarzschild values quoted to four digits. At every later spin, the initial field and frequency are constructed only from the preceding accepted eigenpairs on the same branch. Thus, the solution at one value of $a$ provides the initial guess for the neighboring Kerr problem. When one such initial guess is not sufficiently robust, several independently optimized trial eigenpairs are generated from local extrapolations of the eigenpairs already obtained at the preceding spins of the same branch. Hereafter, a ``candidate'' means one such trial eigenpair at a fixed spin, not a distinct physical mode. The optimization algorithm, convergence criteria, extrapolation rules, and the special high-spin brackets are documented in Appendix~\ref{app:continuation_details}.

\subsection{Spin continuation}
\label{subsec:spin_cont}

We follow each Kerr eigenbranch by numerical continuation in the spin parameter $a$. That is, rather than solving the eigenvalue problem independently at every spin, we continuously deform a known solution as the Kerr background is changed, because this proved to be a much more reliable numerical procedure. At step $k$, the new value of $a$ and the corresponding dimensionless spin parameter $\kappa$ update the operator coefficients given in Appendix~\ref{app:hyperboloidal_operator}; the accepted field and frequency at step $k-1$ initialize the new nonlinear eigenvalue solve (step S1 of Algorithm~\ref{alg:pipeline}). This continuation is important for a non-Hermitian spectrum in which damped branches can approach one another in the complex-frequency plane.

Continuation is performed in $a$ because it is the physical parameter of the family: Schwarzschild is the one limit in which every branch is labeled unambiguously by $\ell$, and each small step lets a branch be followed even where the eigenvalues group together at high spin. This loop is stage~2 to stage~4 of the pipeline in Fig.~\ref{fig:methodology_pipeline}. The new initial guess is first adjusted with the frequency fixed, and the field and frequency are then optimized together. That is, at each new spin the frequency is first held at its inherited value while the network adapts the field to the operator of the new spin; only afterwards is the frequency released and optimized together with the field. Releasing it immediately would let it follow a field that is still adapted to the previous spin, which invites a jump to a neighboring branch. This staged procedure is sometimes called curriculum learning in PINN terminology, but here it is more naturally interpreted as standard parameter continuation through the physical spin variable and uses no supervised frequency targets.

The Kerr bound is $a/M=1$. The discrete values of $a/M$ along the sequences, or the so-called continuation grid, with elements $a_k$, $k=0,\dots,15$, in increasing order, is chosen as a coarse scan at low and moderate spins, followed by a refined scan close to extremality,
\begin{align}
    \mathcal{A}_{\mathrm{cont}}
    &=\{0,0.2,0.4,0.6,0.8,0.9,0.94,0.96,0.97\}
    \nonumber\\
    &\quad\cup\{0.98,0.986,0.992,0.996,0.998,0.999,0.9998\}.
    \label{eq:a_continuation}
\end{align}

The grid in Eq.~\eqref{eq:a_continuation} is equivalent, in terms of the dimensionless hyperboloidal spin parameter $\kappa$, to a sequence approaching $\kappa\to1$ without reaching the extremal limit. This is important because the extremal endpoint changes the character of the radial operator and introduces additional spectral crowding associated with near-horizon, slowly damped branches. The progressively smaller spin increments near $a=M$ therefore serve two purposes: they resolve the increasingly rapid spectral variation and reduce the probability of inadvertently switching between nearby eigenbranches. In the hyperboloidal formulation, the QNM boundary conditions are encoded geometrically at $\sigma=0$ and $\sigma=1$, and the network represents the regular field $h(\sigma,u)$. Continuation is therefore used to preserve branch identity as the operator varies with $\kappa$, rather than to regularize explicit radial prefactors.

At each spin the optimization may be run several times from different starting guesses, and more than one of the resulting fields can satisfy the Teukolsky residual to comparable accuracy while belonging to different QNM branches. A rule is therefore needed to decide which one continues the branch under study (cf.~stage~4 (\textit{select, solve, continue}) of Fig.~\ref{fig:methodology_pipeline} and step S2 of Algorithm~\ref{alg:pipeline}). At moderate spins, the accepted eigenpair directly initializes the next problem. In more crowded parts of the spectrum, several trial eigenpairs are optimized from extrapolations of the accepted same-branch history. Generating several such initial guesses provides a numerical safeguard against the nonlinear solver converging to a nearby branch. At fixed $(\ell,m,n)$ and $a_k$, the $k$-th spin value of the continuation grid in Eq.~\eqref{eq:a_continuation}, candidate $j\in\{1,\ldots,N_k\}$, one of the $N_k$ candidates optimized at that spin,  produces the optimized trial eigenpair $(\omega_k^{(j)},h_k^{(j)})$. Throughout, the subscript $k$ labels the spin and the superscript $(j)$ the candidate computed at that spin, so that $\omega_k^{(j)}$ is the frequency returned by the $j$-th optimization at $a_k$. Once all candidates at $a_{k}$ have been optimized, one of them has to be accepted. That decision is made with a selection score $\mathcal{S}_{k}^{(j)}$: a single real number, computed for each candidate after its optimization has finished, which is smaller the better the candidate continues the mode being followed. It is defined as
\begin{equation}
    \mathcal{S}_{k}^{(j)}
    =\mathcal{S}_{\mathrm{conv},k}^{(j)}
    +\mathcal{S}_{\mathrm{cont},k}^{(j)}
    +\mathcal{S}_{\mathrm{pred},k}^{(j)}
    +\mathcal{S}_{\mathrm{trend},k}^{(j)}
    +\mathcal{S}_{\mathrm{ot},k}^{(j)}.
    \label{eq:continuation_score_overview}
\end{equation}
The five terms in Eq.~\eqref{eq:continuation_score_overview} assess, respectively, residual convergence, continuity of the frequency and two-dimensional field, agreement with same-branch history extrapolations, consistency with the local frequency trend, and overtone-specific branch information. Thus, the score is not an additional physical equation; it is a numerical criterion for deciding which of several already converged eigenpairs is the smooth continuation of the previously accepted branch. Only candidates whose final residual is close to the best one reached at that spin are eligible, and this subset is denoted by $\mathcal{E}_k$. A candidate that continues the branch smoothly but solves the equation poorly is thus discarded before the score is consulted: continuity can never compensate for a bad residual. In the near-extremal tail of the co-rotating branches, a zero-damping consistency check and preferences on the real part and the damping act before the score: inside the low-loss envelope, candidates launched from given predictor sources, or with the larger or the smaller damping, are favored (Appendix~\ref{app:continuation_details}). The selected candidate is
\begin{equation}
    j_k^{\star}
    =\underset{j\in\mathcal{E}_k}{\arg\min}\;\mathcal{S}_k^{(j)},
    \qquad
    (\tilde\omega_k,\tilde h_k)
    =\left(\omega_k^{(j_k^{\star})},h_k^{(j_k^{\star})}\right).
    \label{eq:candidate_acceptance}
\end{equation}

Among the eligible candidates, the selected one is that of smallest score, and its frequency becomes the starting point of the frequency solve of Sec.~\ref{subsubsec:frequency_solve}, whose result is the accepted eigenpair at $a_k$ and initializes the next spin. The expanded score, the definitions of all its components and of $\mathcal{E}_k$, and the numerical weights are given in Appendix~\ref{app:continuation_details}, Eqs.~\eqref{eq:continuation_score}--\eqref{eq:loss_envelope}, and in the released code~\cite{Zenodo_2026}.

\subsection{Near-extremal gauge and frequency solve}
\label{subsec:robust_frequency}

The residual does not always pin the frequency down. Toward extremality, and at some lower spins where the residual has reached its noise level, eigenpairs whose frequencies differ by a few percent reach residuals of the same size, so the point where a stochastic optimizer stops depends on its random path and on rounding errors that differ between processors (Sec.~\ref{subsec:hardware}). A change of gauge makes the near-horizon field easier to represent, and a frequency solve (step S3 of Algorithm~\ref{alg:pipeline}) fixes the frequency by an equation.

\subsubsection{Cauchy-horizon-fixing gauge}
\label{subsubsec:cauchy_gauge}

In the radial-fixing gauge the Cauchy horizon sits at $\sigma_{\rm c}=\kappa^{-2}$, which approaches the event horizon as $a\to M$, and the regular field develops a steep layer at $\sigma=1$. For $m\ge2$ and $a/M\ge0.98$ we therefore solve the problem in the Cauchy-horizon-fixing gauge of Ref.~\cite{Assaad_2025}. The regular field $h(\sigma,u)$ of the radial-fixing gauge and the regular field $g(y,u)$ of the Cauchy-horizon-fixing gauge, written in the compact radial coordinate $y$ of that gauge, are related exactly by
\begin{equation}
    \begin{aligned}
        h(\sigma,u)&=(1-\kappa^2\sigma)^{q}\,g(y,u),\qquad
        y=\frac{(1-\kappa^2)\sigma}{1-\kappa^2\sigma},\\
        q&=2i(1+\kappa^2)\,\bar\omega-1,
    \end{aligned}
    \label{eq:gauge_map}
\end{equation}
where $\bar\omega=r_+\omega$ as in Eq.~\eqref{eq:intro_nonseparated_ansatz} and $1-\kappa^2\sigma=(r-r_-)/r$. The exponent $q$ follows from expressing both regular fields through the same master function $\Psi_s$, each with the height function and rescaling of its own gauge~\cite{Assaad_2025}. The coordinate $y\in[0,1]$ spreads the near-horizon layer over an interval of order one, and the factor $(1-\kappa^2\sigma)^{q}$, which varies rapidly there as $\kappa\to1$, is no longer part of the represented field $g$. The frequency is the same in both gauges. Here the network inputs are $(y,u)$ and $\eta(y)$, and the anchor of Eq.~\eqref{eq:hard_anchor} sits at $y=0.5$. At the first spin in this gauge the field of the preceding spin is mapped by Eq.~\eqref{eq:gauge_map}, refitted, and polished at its frequency, and it starts the frequency solve. The operator coefficients are given in Appendix~\ref{app:hyperboloidal_operator}.

\subsubsection{Frequency solve by variable projection}
\label{subsubsec:frequency_solve}

The frequency solve is a numerical polish of the frequency found by training: it starts from the frequency of the selected candidate (or the extrapolated one, Algorithm~\ref{alg:pipeline}), which already lies on the intended branch, and refines it by an equation. It consists of two nested iterations. The \textit{outer iteration} updates the two frequency components. Inside each outer iteration, an \textit{inner iteration}, the field polish, corrects the network weights at the current frequency. Both iterations are of Gauss--Newton type: the residual is replaced by its linearization in the quantities that the iteration corrects (the weights in the inner iteration, the two frequency components in the outer one), and the correction minimizes the norm of that linear model, with a damping term in the inner iteration. Let $\bm{r}(\Theta,\bm{\omega})\in\mathbb{R}^{2N}$ collect the real and imaginary parts of the residual $\mathcal{R}_{\Theta}$ of Eq.~\eqref{eq:pinn_residual} (or of its counterpart in the Cauchy-horizon-fixing gauge), divided by $\sqrt{N}$, on a uniform mesh of $N=65\times65$ points that includes the boundaries; $\Theta$ are the network weights of Eq.~\eqref{eq:nn_ansatz} and $\bm{\omega}=(\omega_R,\omega_I)$, with Jacobians $J_\Theta=\partial\bm{r}/\partial\Theta$ and $J_\omega=\partial\bm{r}/\partial\bm{\omega}$.

The \textit{inner iteration} holds the frequency fixed and corrects the weights by at most $30$ Levenberg--Marquardt steps~\cite{Levenberg_1944,Marquardt_1963},
\begin{equation}
    \left(J_\Theta^{\mathsf T}J_\Theta+\mu\,I\right)\Delta\Theta=-J_\Theta^{\mathsf T}\bm{r}.
    \label{eq:lm_polish}
\end{equation}
For $\mu=0$, this is the Gauss--Newton step; a larger $\mu$ shortens the step toward the gradient direction, which keeps it stable when $J_\Theta^{\mathsf T}J_\Theta$ is singular or nearly so, as it is for a network with thousands of weights. The parameter $\mu$ is adapted at each step, and the linear systems are solved by conjugate gradients from Jacobian--vector products. It runs in double precision from the field accepted at the previous spin (at $a/M=0$, the selected candidate). We write $\hat\Theta(\bm{\omega})$ for the weights it reaches at frequency $\bm\omega$, so $h_{\hat\Theta(\bm\omega)}$ is the polished field. The $3\times56$ and $4\times84$ networks have more weights ($29{,}742$ and $87{,}198$) than the mesh has residual values ($8{,}450$), so many weight vectors fit the mesh equally well; the polish selects the one it reaches from its starting field.

The \textit{outer iteration} chooses the frequency that minimizes the polished residual $\varphi(\bm{\omega})=\tfrac12\|\bm{r}(\hat\Theta(\bm{\omega}),\bm{\omega})\|^{2}$, a function of two unknowns only. Linearizing in $\bm\omega$ gives the Gauss--Newton step that minimizes $\|\bm r+J_\omega\Delta\bm\omega\|$, but it ignores that the polished field itself changes with the frequency. Variable projection~\cite{Golub_Pereyra_1973,Golub_Pereyra_2003} eliminates parameters that enter the model linearly; in the same spirit, we account to first order for this response in the weights $\Theta_{\rm out}$ of the output layer, which maps the last hidden features onto the two field components. For a hidden width $d$ this layer has $6d+6$ weights, from $198$ for $2\times32$ to $510$ for $4\times84$. They are fewer than the residual values, so the projection cannot remove the whole residual, and since that layer is quadratic in its weights, the elimination holds only to first order. The residual changes that a small readjustment of these weights could produce span the column space of $J_{\rm out}=\partial\bm{r}/\partial\Theta_{\rm out}$. With the singular-value decomposition $J_{\rm out}=U\Sigma V^{\mathsf T}$, the orthogonal projector onto that space is
\begin{equation}
    P=\tilde U\tilde U^{\mathsf T},
    \label{eq:varpro_projector}
\end{equation}
where the columns of $\tilde U$ are the left singular vectors whose singular values exceed $10^{-10}$ times the largest; smaller ones set no reliable direction. Thus $P^2=P=P^{\mathsf T}$, and the factor $I-P$ is applied to $\bm r$ and $J_\omega$ before the step is computed; it keeps only the residual that the output layer cannot cancel, which is the part that really needs the frequency to change. The step is the least-squares solution of $(I-P)(\bm r+J_\omega\Delta\bm\omega)\approx0$,
\begin{equation}
    \Delta\bm{\omega}=-\left[J_\omega^{\mathsf T}(I-P)J_\omega\right]^{-1}J_\omega^{\mathsf T}(I-P)\,\bm{r}.
    \label{eq:varpro_step}
\end{equation}
After each frequency step the field is polished again from where it stopped. A step that would change either frequency component by more than $5\%$ is scaled down as a whole, and the iteration ends when both relative changes fall below $10^{-4}$. At a fixed point of the iteration ($\Delta\bm\omega=0$), $J_\omega^{\mathsf T}(I-P)\bm r=0$, which is the stationarity condition $\nabla\varphi=J_\omega^{\mathsf T}\bm r=0$ when the polished field is stationary in the weights, $J_\Theta^{\mathsf T}\bm r=0$ (in particular $J_{\rm out}^{\mathsf T}\bm r=0$, so that $P\bm r=0$). The iteration converged at all 336 reported points, in at most $32$ steps (if it did not, a profile of the residual at fixed trial frequencies would replace it); Appendix~\ref{app:frequency_solve} shows how the error of the frequency changes over the steps. Steps S4--S6 of Algorithm~\ref{alg:pipeline} follow, with the settings of Table~\ref{tab:app_common_settings}: training with $\omega$ fixed at the solution (S4); for the overtones at $0.2\le a/M\le0.96$, double-precision limited-memory Broyden--Fletcher--Goldfarb--Shanno (L-BFGS) steps~\cite{Nocedal_1980,Liu_Nocedal_1989} on the weights and the frequency together (S5), minimizing the loss of Eq.~\eqref{eq:total_loss} without the frequency-control, overtone branch, and residual-gradient terms and kept only if this objective also decreases on an independent mesh (kept at all 42 points, with frequency changes of at most $0.14\%$ over the three networks); and a Levenberg--Marquardt polish of the field at fixed frequency (S6), started from whichever of the trained field and the last field of the solve has the smaller residual on an offset mesh.

Apart from S5, the frequency is thus fixed by the stationarity of the plain residual after a deterministic, warm-started polish of bounded depth rather than by where a stochastic optimization stopped, and auxiliary loss terms enter it only through the starting field and frequency. Since the two larger networks have more weights than the mesh has residual values, the polished residual, and hence its stationary point, depend on the warm start and the bounded depth of the polish; the residual on a second mesh (Appendix~\ref{app:frequency_solve}) shows that the accepted fields do not merely interpolate the solve mesh. In earlier test runs (not those of Appendix~\ref{app:frequency_solve}), locating the frequency from residuals at fixed trial frequencies, with the field optimized at each, missed the $(2,2,0)$ damping by $6.4\%$ on the $4\times84$ network at $a/M=0.9$, where the four candidates spread over $4\%$ in the real part, and the $(3,3,0)$ damping by $1.25\%$ on the $2\times32$ network at $a/M=0$; started from the same fields, the solve converged in four and three steps to within $10^{-4}\%$ of the reference.

\subsection{Physical equations and numerical controls}
\label{subsubsec:phys_eqs}

Only $\mathcal{R}_{\Theta}=0$ in Eq.~\eqref{eq:pinn_residual}, together with regularity at the hyperboloidal boundaries and at the axis, defines the physical eigenvalue problem. The anchor in Eq.~\eqref{eq:hard_anchor} fixes the arbitrary complex normalization. The relative residual and the weak mode, overtone, and frequency-control terms are auxiliary devices for nonlinear optimization and local branch tracking. They neither change the Teukolsky operator nor supply additional boundary data. The curvature, overtone, and frequency-control terms are not zero at the exact eigenfunction, so they displace the minimum of the total loss. The displacement scales with their weights, of order $10^{-5}$ to $10^{-4}$ against the unit weight of the residual for the mode and overtone terms (the frequency-control term is normalized to the residual but acts only in step S1), and they serve only to choose the basin: the frequency solve uses the plain residual, not the penalized objective (only in step S5 do they shift the overtone frequencies, by at most $0.14\%$ on the three networks and $0.03\%$ on the $4\times84$ network). Whatever bias survives is bounded by the Leaver comparison of Sec.~\ref{sec:Results}, where every component error stays below $0.23\%$.

Likewise, the selection score does not define an eigenmode. It chooses among separately optimized trial solutions that already have comparable residual quality. Its continuity checks reduce unintended jumps between nearby branches as the spin changes.

The independent Leaver-based reference spectrum is absent from the loss, the continuation proposals, the convergence tests, the frequency solve, and the selection score; in the final calculations it is used only for the assessment in Sec.~\ref{sec:Results}, and at $a/M=0$ Schwarzschild values or eikonal estimates seed the first optimization. The reference values were, however, used while the method was developed on these Kerr branches, to find the spins at which a version of the method failed (as in the test runs of Sec.~\ref{subsubsec:frequency_solve}). Appendix~\ref{app:continuation_details} reports the complete implementation and numerical settings.

\section{Results}
\label{sec:Results}

For the benchmarking of our results, the frequencies used for comparison are computed with the Python package \texttt{qnm}~\cite{Stein_2019}, which implements the Leaver continued-fraction method and a spectral treatment of the angular sector. The Kerr QNM spectrum has no closed form, and continued-fraction values are among the most accurate available; their precision far exceeds the deviations reported below, so we call them the reference and treat them as exact for the present comparison.
%
%
This comparison is the assessment block displayed at the last step of  Fig.~\ref{fig:methodology_pipeline}, and it is performed only after each continuation sequence is complete. Because the method was developed on these branches (Sec.~\ref{subsubsec:phys_eqs}), the comparison validates the final method rather than serving as a blind test. For either frequency component $X\in\{\Re(\omega),\Im(\omega)\}$, we report
\begin{equation}
    \epsilon_X=100\left|\frac{X_{\mathrm{PINN}}-X_{\mathrm{ref}}}{X_{\mathrm{ref}}}\right|\,\%.
    \label{eq:component_error}
\end{equation}

\begin{figure*}[!t]
    \centering
    \includegraphics[width=\textwidth]{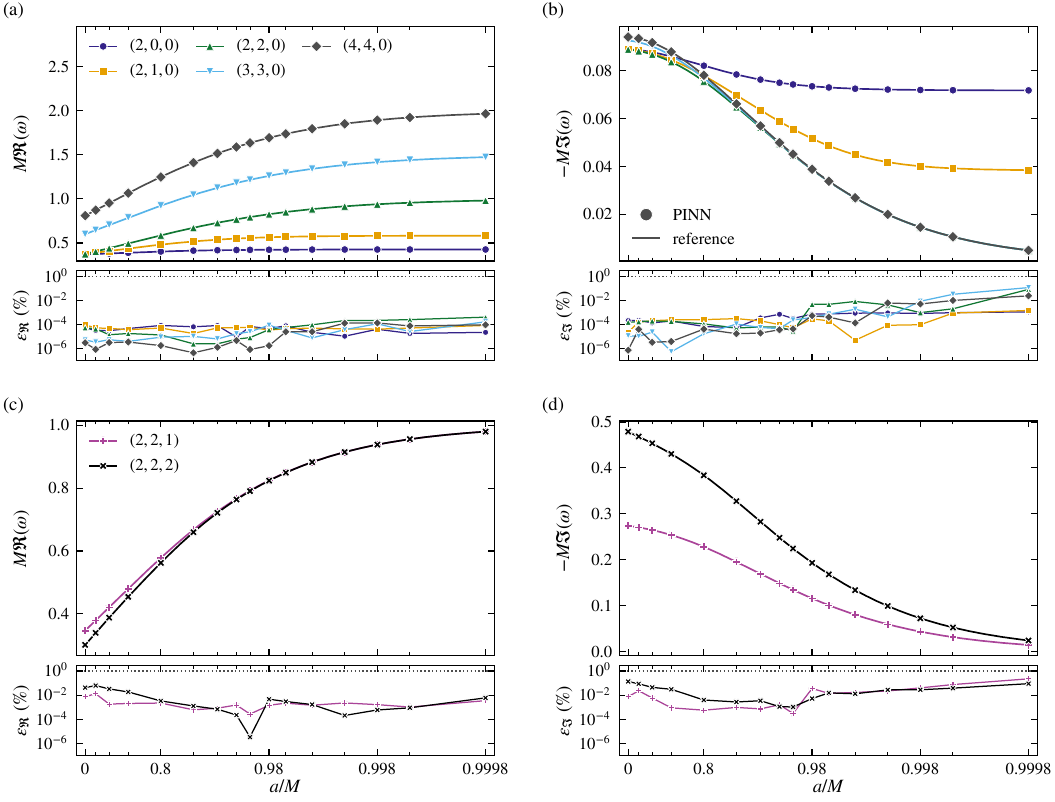}
    \caption{\justifying Real frequency [(a), (c)] and damping rate [(b), (d)] of the fundamental modes (top) and of the overtones (bottom) for the $4\times84$ network. Markers are PINN values at the 16 trained spins and curves the reference values; the strips give the relative errors $\epsilon_{\Re}$ and $\epsilon_{\Im}$ of Eq.~\eqref{eq:component_error}, with the $1\%$ level dotted. The spin axis is logarithmic in $1-a/M$.}
    \label{fig:frequency_progression}
\end{figure*}

\subsection{Kerr frequency spectra}

The set of modes considered spans the observationally relevant branches and the numerically demanding ones. The fundamental quadrupole $(2,2,0)$ dominates comparable-mass ringdowns and is the branch that any solver must reproduce. Excitation of the leading subdominant harmonics is governed by the progenitor mass ratio: odd-$m$ multipoles such as $(3,3,0)$ vanish in the equal-mass nonspinning limit, whereas $(4,4,0)$ is excited with a non-negligible amplitude~\cite{Kamaretsos_2012}. Every additional resolved mode constrains the remnant mass and spin independently, which is the basis of multi-mode spectroscopy~\cite{Dreyer_2004}. The first overtone $(2,2,1)$ provides the same kind of constraint within a single angular harmonic, and it is expected to be excited with a non-negligible amplitude. The second overtone $(2,2,2)$ shares the same angular structure but decays faster still, and it is included mainly as a harder numerical test. Unlike the frequencies, the mode amplitudes and phases depend on the properties of the progenitor binary and must therefore be determined from numerical relativity.

Apart from observational relevance, an important test for every code is the accurate calculation of the most numerically challenging QNMs. Co-rotating $m=\ell$ branches become zero-damped toward extremality, so the eigenvalue approaches the real axis and the eigenfunction steepens at the horizon~\cite{Yang_2013,Yang_2013_extended}, whereas $(2,2,1)$ and $(2,2,2)$ remain spectrally adjacent to $(2,2,0)$ throughout the continuation. The low-$m$ branches $(2,0,0)$ and $(2,1,0)$ stay damped and provide the complementary control case. 

The calculated mode frequencies as functions of $a/M$ are displayed in Fig.~\ref{fig:frequency_progression} for all seven considered modes, where also the reference Leaver dependencies are plotted with lines. The horizontal coordinate is logarithmic in the dimensionless distance to extremality, $1-a/M$, so the densely sampled high-spin tail remains visible. The $m=0$ branch changes comparatively slowly, whereas the co-rotating branches increase rapidly in real frequency and become less damped as $a\to M$. The PINN markers remain visually coincident with the reference curves over the full spin range, including the two overtones. Across all reported frequencies,  every real- and imaginary-part error falls below $0.23\%$ for the production $4\times84$ network, and every real part agrees within $0.063\%$. The largest mean component error of any mode is $0.033\%$, for the damping of $(2,2,2)$, and the maximum component error at the final spin $a/M=0.9998$ is $0.229\%$, for the damping of $(2,2,1)$.

\begin{table*}[!t]
\centering
\scriptsize
\setlength{\tabcolsep}{2.0pt}
\renewcommand{\arraystretch}{1.20}
\caption{PINN frequencies of the $4\times84$ network at every spin, with their relative errors of Eq.~\eqref{eq:component_error} in percent: (a) real part $M\Re(\omega)$ and $\epsilon_{\Re}$; (b) imaginary part $M\Im(\omega)$ and $\epsilon_{\Im}$. Errors below $5\times10^{-4}\%$ are given as $<0.001$.}
\label{tab:frequency}
\textit{(a) Real part}\par\smallskip
{\begin{tabular*}{\textwidth}{@{\extracolsep{\fill}}c*{7}{rr}@{}}
\hline\hline
$a/M$ & \multicolumn{2}{c}{$(2,0,0)$} & \multicolumn{2}{c}{$(2,1,0)$} & \multicolumn{2}{c}{$(2,2,0)$} & \multicolumn{2}{c}{$(3,3,0)$} & \multicolumn{2}{c}{$(4,4,0)$} & \multicolumn{2}{c}{$(2,2,1)$} & \multicolumn{2}{c}{$(2,2,2)$} \\
 & PINN & $\epsilon_{\Re}$ & PINN & $\epsilon_{\Re}$ & PINN & $\epsilon_{\Re}$ & PINN & $\epsilon_{\Re}$ & PINN & $\epsilon_{\Re}$ & PINN & $\epsilon_{\Re}$ & PINN & $\epsilon_{\Re}$ \\
\hline
0.0000 & 0.373672 & $<0.001$ & 0.373672 & $<0.001$ & 0.373672 & $<0.001$ & 0.599443 & $<0.001$ & 0.809178 & $<0.001$ & 0.346685 & 0.007 & 0.301181 & 0.042 \\
0.2000 & 0.375124 & $<0.001$ & 0.388248 & $<0.001$ & 0.402146 & $<0.001$ & 0.644787 & $<0.001$ & 0.871718 & $<0.001$ & 0.379027 & 0.013 & 0.339498 & 0.063 \\
0.4000 & 0.379682 & $<0.001$ & 0.407979 & $<0.001$ & 0.439842 & $<0.001$ & 0.703650 & $<0.001$ & 0.952500 & $<0.001$ & 0.420854 & 0.002 & 0.387908 & 0.033 \\
0.6000 & 0.388054 & $<0.001$ & 0.435969 & $<0.001$ & 0.494045 & $<0.001$ & 0.786223 & $<0.001$ & 1.064981 & $<0.001$ & 0.479816 & 0.002 & 0.454095 & 0.018 \\
0.8000 & 0.401918 & $<0.001$ & 0.480231 & $<0.001$ & 0.586017 & $<0.001$ & 0.921885 & $<0.001$ & 1.247547 & $<0.001$ & 0.577935 & 0.002 & 0.562221 & 0.003 \\
0.9000 & 0.412005 & $<0.001$ & 0.516291 & $<0.001$ & 0.671614 & $<0.001$ & 1.044637 & $<0.001$ & 1.410416 & $<0.001$ & 0.667662 & 0.001 & 0.659818 & 0.001 \\
0.9400 & 0.416862 & $<0.001$ & 0.536674 & $<0.001$ & 0.727875 & $<0.001$ & 1.124097 & $<0.001$ & 1.514785 & $<0.001$ & 0.725702 & 0.001 & 0.721501 & 0.001 \\
0.9600 & 0.419491 & $<0.001$ & 0.549213 & $<0.001$ & 0.767674 & $<0.001$ & 1.179861 & $<0.001$ & 1.587589 & $<0.001$ & 0.766337 & 0.002 & 0.763832 & $<0.001$ \\
0.9700 & 0.420856 & $<0.001$ & 0.556329 & $<0.001$ & 0.793208 & $<0.001$ & 1.215469 & $<0.001$ & 1.633903 & $<0.001$ & 0.792274 & $<0.001$ & 0.790516 & $<0.001$ \\
0.9800 & 0.422254 & $<0.001$ & 0.564155 & $<0.001$ & 0.825430 & $<0.001$ & 1.260228 & $<0.001$ & 1.691943 & $<0.001$ & 0.824864 & 0.001 & 0.823817 & 0.005 \\
0.9860 & 0.423109 & $<0.001$ & 0.569227 & $<0.001$ & 0.850272 & $<0.001$ & 1.294616 & $<0.001$ & 1.736405 & $<0.001$ & 0.849915 & 0.002 & 0.849214 & 0.003 \\
0.9920 & 0.423974 & $<0.001$ & 0.574536 & $<0.001$ & 0.883161 & $<0.001$ & 1.339991 & $<0.001$ & 1.794920 & $<0.001$ & 0.882990 & 0.001 & 0.882626 & 0.002 \\
0.9960 & 0.424557 & $<0.001$ & 0.578085 & $<0.001$ & 0.914900 & $<0.001$ & 1.383633 & $<0.001$ & 1.851039 & $<0.001$ & 0.914854 & 0.002 & 0.914686 & $<0.001$ \\
0.9980 & 0.424851 & $<0.001$ & 0.579796 & $<0.001$ & 0.938521 & $<0.001$ & 1.416028 & $<0.001$ & 1.892602 & $<0.001$ & 0.938517 & 0.002 & 0.938448 & 0.001 \\
0.9990 & 0.424998 & $<0.001$ & 0.580624 & $<0.001$ & 0.955852 & $<0.001$ & 1.439748 & $<0.001$ & 1.922994 & $<0.001$ & 0.955857 & 0.001 & 0.955842 & 0.001 \\
0.9998 & 0.425116 & $<0.001$ & 0.581273 & $<0.001$ & 0.979837 & $<0.001$ & 1.472513 & $<0.001$ & 1.964905 & $<0.001$ & 0.979874 & 0.004 & 0.979896 & 0.006 \\
\hline\hline
\end{tabular*}
}
\par\bigskip
\textit{(b) Imaginary part}\par\smallskip
{\begin{tabular*}{\textwidth}{@{\extracolsep{\fill}}c*{7}{rr}@{}}
\hline\hline
$a/M$ & \multicolumn{2}{c}{$(2,0,0)$} & \multicolumn{2}{c}{$(2,1,0)$} & \multicolumn{2}{c}{$(2,2,0)$} & \multicolumn{2}{c}{$(3,3,0)$} & \multicolumn{2}{c}{$(4,4,0)$} & \multicolumn{2}{c}{$(2,2,1)$} & \multicolumn{2}{c}{$(2,2,2)$} \\
 & PINN & $\epsilon_{\Im}$ & PINN & $\epsilon_{\Im}$ & PINN & $\epsilon_{\Im}$ & PINN & $\epsilon_{\Im}$ & PINN & $\epsilon_{\Im}$ & PINN & $\epsilon_{\Im}$ & PINN & $\epsilon_{\Im}$ \\
\hline
0.0000 & -0.088962 & $<0.001$ & -0.088962 & $<0.001$ & -0.088962 & $<0.001$ & -0.092703 & $<0.001$ & -0.094164 & $<0.001$ & -0.273894 & 0.007 & -0.478932 & 0.137 \\
0.2000 & -0.088700 & $<0.001$ & -0.088488 & $<0.001$ & -0.088311 & $<0.001$ & -0.091973 & $<0.001$ & -0.093537 & $<0.001$ & -0.270609 & 0.023 & -0.468283 & 0.088 \\
0.4000 & -0.087826 & $<0.001$ & -0.087257 & $<0.001$ & -0.086882 & $<0.001$ & -0.090218 & $<0.001$ & -0.091829 & $<0.001$ & -0.264749 & 0.006 & -0.453462 & 0.045 \\
0.6000 & -0.085994 & $<0.001$ & -0.084564 & $<0.001$ & -0.083765 & $<0.001$ & -0.086385 & $<0.001$ & -0.087909 & $<0.001$ & -0.253849 & 0.001 & -0.430448 & 0.031 \\
0.8000 & -0.082156 & $<0.001$ & -0.077955 & $<0.001$ & -0.075629 & $<0.001$ & -0.076995 & $<0.001$ & -0.078126 & $<0.001$ & -0.228150 & 0.001 & -0.383910 & 0.004 \\
0.9000 & -0.078483 & $<0.001$ & -0.069804 & $<0.001$ & -0.064869 & $<0.001$ & -0.065463 & $<0.001$ & -0.066156 & $<0.001$ & -0.195254 & 0.001 & -0.327527 & 0.003 \\
0.9400 & -0.076330 & $<0.001$ & -0.063549 & $<0.001$ & -0.056231 & $<0.001$ & -0.056537 & $<0.001$ & -0.056982 & $<0.001$ & -0.169018 & 0.001 & -0.282817 & 0.003 \\
0.9600 & -0.075028 & 0.001 & -0.058791 & $<0.001$ & -0.049434 & $<0.001$ & -0.049609 & $<0.001$ & -0.049911 & $<0.001$ & -0.148455 & 0.002 & -0.248005 & 0.001 \\
0.9700 & -0.074307 & $<0.001$ & -0.055644 & $<0.001$ & -0.044790 & $<0.001$ & -0.044904 & $<0.001$ & -0.045132 & $<0.001$ & -0.134453 & $<0.001$ & -0.224408 & 0.001 \\
0.9800 & -0.073533 & 0.001 & -0.051643 & $<0.001$ & -0.038628 & 0.005 & -0.038688 & $<0.001$ & -0.038837 & 0.001 & -0.115959 & 0.036 & -0.193297 & 0.005 \\
0.9860 & -0.073041 & 0.001 & -0.048614 & $<0.001$ & -0.033657 & 0.005 & -0.033688 & 0.001 & -0.033790 & $<0.001$ & -0.100971 & 0.014 & -0.168314 & 0.016 \\
0.9920 & -0.072528 & 0.001 & -0.044856 & $<0.001$ & -0.026789 & 0.008 & -0.026797 & 0.002 & -0.026854 & $<0.001$ & -0.080363 & 0.017 & -0.133947 & 0.013 \\
0.9960 & -0.072173 & 0.001 & -0.041794 & $<0.001$ & -0.019865 & 0.004 & -0.019865 & $<0.001$ & -0.019891 & 0.006 & -0.059591 & 0.022 & -0.099324 & 0.028 \\
0.9980 & -0.071991 & 0.001 & -0.040078 & $<0.001$ & -0.014534 & 0.001 & -0.014534 & 0.009 & -0.014546 & 0.005 & -0.043587 & 0.040 & -0.072664 & 0.028 \\
0.9990 & -0.071899 & 0.001 & -0.039179 & 0.001 & -0.010530 & 0.002 & -0.010533 & 0.033 & -0.010537 & 0.010 & -0.031568 & 0.076 & -0.052635 & 0.039 \\
0.9998 & -0.071826 & 0.001 & -0.038443 & 0.002 & -0.004863 & 0.085 & -0.004873 & 0.120 & -0.004867 & 0.025 & -0.014568 & 0.229 & -0.024314 & 0.090 \\
\hline\hline
\end{tabular*}
}
\end{table*}

The complete numerical landscape is reported in Table~\ref{tab:frequency}, which lists the PINN value and its reference-based component error at every spin for every mode. In particular, at $a/M=0.9998$ the production PINN gives $M\omega=0.979837-0.004863i$ for $(2,2,0)$, $M\omega=0.979874-0.014568i$ for $(2,2,1)$, and $M\omega=0.979896-0.024314i$ for $(2,2,2)$, illustrating that the continuation and overtone controls distinguish three nearby real-frequency branches with substantially different damping rates. The same three branches are visible in Fig.~\ref{fig:frequency_progression}, where the overtone markers in Fig.~\ref{fig:frequency_progression}(c) track the $(2,2,0)$ real frequency of Fig.~\ref{fig:frequency_progression}(a) while the three damping rates stay clearly separated.

The largest real-frequency error in Table~\ref{tab:frequency}(a) occurs for $(2,2,2)$ at $a/M=0.2$, while the largest damping-rate error in Table~\ref{tab:frequency}(b) occurs for $(2,2,1)$ at $a/M=0.9998$. For the co-rotating fundamental modes the damping error grows toward extremality, from $10^{-3}\%$ or less at moderate spins to between $0.025\%$ and $0.120\%$ at $a/M=0.9998$ ($0.229\%$ and $0.090\%$ for the two overtones), where the damping rate has fallen to about $5\%$ of its Schwarzschild value; the real-part errors grow too but stay below $5\times10^{-4}\%$ for the fundamental modes. In Fig.~\ref{fig:frequency_progression} the markers nevertheless remain on the reference curves at the densest near-extremal spins. At $a/M=0.9998$, the damping magnitudes of the co-rotating $(2,2,0)$, $(3,3,0)$, and $(4,4,0)$ fundamental QNMs are all close to $4.9\times10^{-3}$, while their real frequencies increase with $m$. The $(2,2,1)$ and $(2,2,2)$ real frequencies lie within $6\times10^{-5}$ of that of $(2,2,0)$, while their damping magnitudes are $3.0$ and $5.0$ times larger, the ratios $(n+1/2)/(1/2)$ expected for zero-damped overtones (Sec.~\ref{subsec:near_extremal_scaling}). The two components therefore retain enough information to distinguish the overtones from the neighboring fundamental branch.

\subsection{Near-extremal zero-damping scaling}
\label{subsec:near_extremal_scaling}

\begin{figure*}[!t]
    \centering
    \includegraphics[width=\textwidth]{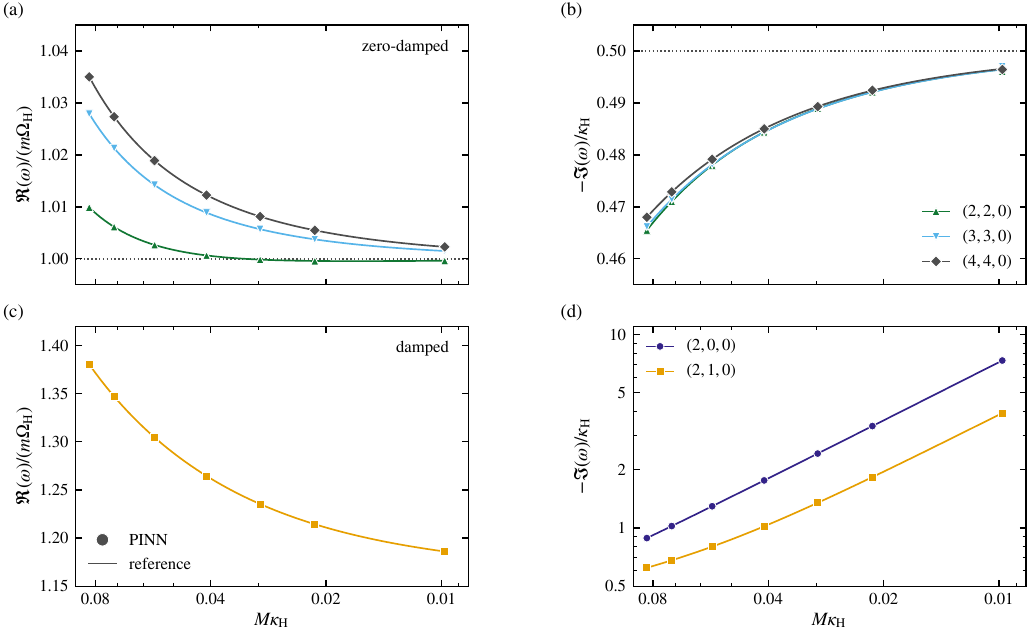}
    \caption{\justifying Near-extremal scaling of the fundamental modes for $0.98\le a/M\le0.9998$. (a), (c) Real frequency over $m\Omega_{\rm H}$ (only $m>0$); (b), (d) damping rate over $\kappa_{\rm H}$. Top: zero-damped branches; bottom: damped branches. Dotted lines mark $\Re(\omega)=m\Omega_{\rm H}$ and $-\Im(\omega)=\kappa_{\rm H}/2$. Markers are PINN values ($4\times84$) and curves reference values.}
    \label{fig:near_extremal_scaling}
\end{figure*}

The approach to extremality becomes more transparent when the QNM frequencies are measured against the natural horizon scales. With $r_{\pm}=M\pm\sqrt{M^2-a^2}$, the horizon angular velocity and surface gravity are $\Omega_{\rm H}=a/(r_+^2+a^2)$ and $\kappa_{\rm H}=(r_+-r_-)/[2(r_+^2+a^2)]$, respectively. The angular velocity sets the rotation frequency of the horizon, whereas the surface gravity sets the near-horizon redshift and decay scale. The extremal limit $a\rightarrow M$ therefore has finite $\Omega_{\rm H}$ but $M\kappa_{\rm H}\rightarrow0$. Figure~\ref{fig:near_extremal_scaling} uses the seven spin values $0.98\le a/M\le0.9998$ to test both the damping scaling and the locking of the oscillation frequency to the horizon rotation.

The nearly extremal Kerr spectrum can contain two physically distinct families~\cite{Yang_2013,Yang_2013_extended}, both of which appear in Fig.~\ref{fig:near_extremal_scaling}: Figs.~\ref{fig:near_extremal_scaling}(a) and \ref{fig:near_extremal_scaling}(b) collect the zero-damped branches and Figs.~\ref{fig:near_extremal_scaling}(c) and \ref{fig:near_extremal_scaling}(d) the damped ones. Zero-damped modes are supported by the near-horizon region: their damping rate vanishes and their real frequency approaches the horizon-rotation frequency as extremality is approached. Damped modes remain associated with an exterior potential barrier and retain a finite decay rate in the same limit. When both families coexist in one $(\ell,m)$ sector, the subextremal overtone sequence bifurcates as the spin increases. The continuation label $n$ alone then does not identify the limiting family, which makes branch tracking necessary.

For the co-rotating zero-damped family, the leading near-extremal asymptotics, derived analytically in Refs.~\cite{Hod_2008,Hod_2009}, can be written as
\begin{equation}
    \Re(\omega_{\ell m n})=m\Omega_{\rm H}+\mathcal{O}(\kappa_{\rm H}),
    \qquad
    -\Im(\omega_{\ell m n})\simeq\left(n+\tfrac{1}{2}\right)\kappa_{\rm H}.
    \label{eq:zero_damping_scaling}
\end{equation}

Thus, $-\Im(\omega)/\kappa_{\rm H}\rightarrow n+1/2$. The value $1/2$ is specifically the leading prediction for the fundamental zero-damped branch, $n=0$, rather than a fitted constant or a universal value for every Kerr QNM. A zero-damped overtone with index $n>0$ instead approaches the corresponding half-integer $n+1/2$. The two overtones computed here follow this limit: at $a/M=0.9998$ the PINN values give $-\Im(\omega)/\kappa_{\rm H}=1.49$ for $(2,2,1)$ and $2.48$ for $(2,2,2)$. Their tail starts use predictors built on this asymptote (Appendix~\ref{app:continuation_details}), but the frequency solve sets the accepted values.  In Fig.~\ref{fig:near_extremal_scaling} the $n=0$ limit is the dotted guide that the zero-damped branches approach from below in panel (b) as $M\kappa_{\rm H}$ decreases.

The effective-potential picture explains why the branches separate in this way~\cite{Yang_2013,Yang_2013_extended}. For sufficiently corotating modes, the relevant potential peak approaches the horizon and its instability timescale becomes proportional to $\kappa_{\rm H}^{-1}$. In the eikonal description, this behavior is governed by $\mu=m/(\ell+1/2)$, with a transition near $\mu_{\rm c}\simeq0.74$. The $(2,2,0)$, $(3,3,0)$, and $(4,4,0)$ modes have $\mu=0.80$, $0.86$, and $0.89$, respectively, and follow the zero-damped family. At $a/M=0.9998$, the PINN values give $-\Im(\omega)/\kappa_{\rm H}=0.496$, $0.497$, and $0.496$. Their real-frequency ratios are $\Re(\omega)/(m\Omega_{\rm H})=1.000$, $1.002$, and $1.002$, so both components satisfy the expected near-horizon scaling. These three branches lie on the dotted guides in Figs.~\ref{fig:near_extremal_scaling}(a) and \ref{fig:near_extremal_scaling}(b).

The lower-$\mu$ sectors can retain an exterior potential maximum and therefore support damped modes in addition to zero-damped modes (for $m=0$, only damped modes). The continuation used here follows the damped $(2,0,0)$ and $(2,1,0)$ branches connected to the corresponding low-spin fundamental modes. Their damping rates remain finite over the sampled interval; consequently, division by $\kappa_{\rm H}\rightarrow0$ raises $-\Im(\omega)/\kappa_{\rm H}$ to $7.33$ and $3.92$, respectively, at the final spin. This is the steep rise in Fig.~\ref{fig:near_extremal_scaling}(d). This growth is the expected signature of a damped branch, not a loss of numerical accuracy. The $(2,1,0)$ real-frequency ratio also reaches $1.186$ at the final spin, well away from 1, while the $m=0$ branch has no nonzero horizon-locking frequency for the quotient of Figs.~\ref{fig:near_extremal_scaling}(a) and \ref{fig:near_extremal_scaling}(c).

The distinction affects both the dynamics and the numerical spectrum. The decay time of a zero-damped mode grows as $\kappa_{\rm H}^{-1}$, its quality factor increases, and its frequency joins an increasingly crowded set near $m\Omega_{\rm H}$. A superposition of these long-lived modes can produce an early power-law ringdown before the late-time exponential decay~\cite{Yang_2013_extended}. Damped modes instead retain finite decay times and can provide a separate, more rapidly decaying contribution when the two families coexist. Excitation amplitudes still determine whether either contribution is observable, but the spectral crowding alone makes continuation and branch diagnostics increasingly important near extremality. The recovered spectra distinguish these two limiting behaviors without using the reference frequencies during training or branch selection.

\begin{figure*}[!t]
    \centering
    \includegraphics[width=\textwidth]{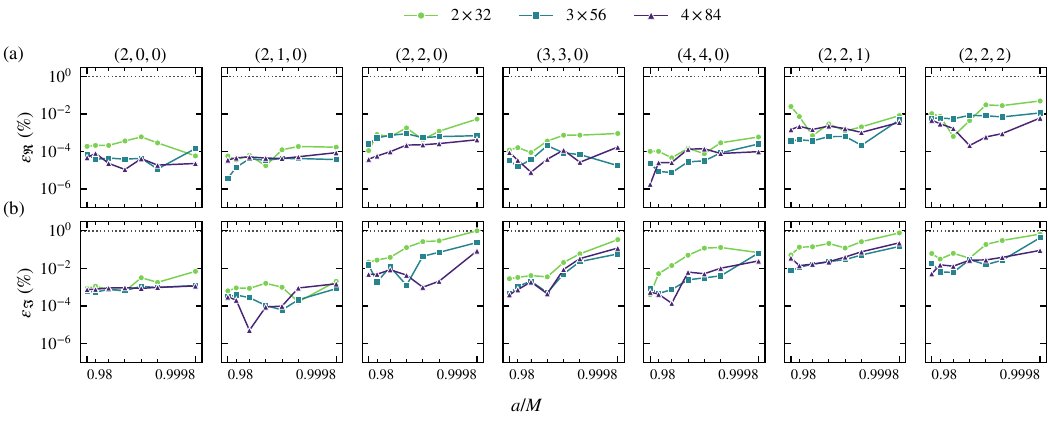}
    \caption{\justifying Relative errors of (a) the real part and (b) the damping rate for $0.98\le a/M\le0.9998$, for the three networks and the seven modes. The dotted line marks $1\%$.}
    \label{fig:architecture_comparison}
\end{figure*}

\subsection{Numerical robustness under changes of network size}
\label{subsec:numerical_robustness}

To test sensitivity to the numerical trial-space resolution, we repeat the complete seven-branch, 16-spin calculation with three network sizes while keeping the physical formulation and continuation protocol fixed.

The curves in Fig.~\ref{fig:architecture_comparison} show the reference-based real- and imaginary-frequency errors in the near-extremal interval $0.98\leq a/M\leq0.9998$ for all three architectures and all modes, while Table~\ref{tab:architecture_results} summarizes the maximum errors over the complete spin grid and the computational cost. The $4\times84$ network is the production baseline used for the principal frequency results, tables, and figures. For this production resolution, the errors of both frequency components remain below $0.23\%$ for every retained mode; the worst component error decreases from $0.427\%$ with the $3\times56$ representation to $0.229\%$ with the $4\times84$ representation. The width, the depth, and the activation functions of a neural trial space are discretization choices that must be fixed before any training begins, and no single architecture is optimal for every problem, so the three networks compared here are prescribed a priori rather than obtained from an open search for an optimum. Within this set, the smallest network lies above the two larger ones for most branches at the highest spins, as Fig.~\ref{fig:architecture_comparison} shows, whereas $3\times56$ and $4\times84$ show no systematic ordering; such an ordering is an empirical observation for the present problem and cannot be expected to be a general property of arbitrary larger networks. The $2\times32$ results satisfy the $1\%$ threshold for every branch, but come closest to it, with $0.997\%$ in the $(2,2,0)$ damping at $a/M=0.9998$ (Sec.~\ref{subsec:hardware}). Panel (b) of Table~\ref{tab:architecture_results} reports the cost of a common residual-evaluation workload, for one training step and one inference pass. The timing protocol and hardware are specified in Appendix~\ref{app:continuation_details}; these measurements do not enter the physical comparison.

\begin{table}[!t]
\centering
\small
\setlength{\tabcolsep}{2pt}
\caption{(a) Maximum relative errors, in percent, over the spin grid for the three networks; values below $5\times10^{-4}\%$ are given as $<0.001$. (b) Central processing unit (CPU) time of one training step and of one inference pass on a common 4096-point workload.}
\label{tab:architecture_results}
\textit{(a) Maximum error across the spin grid}\par\smallskip
{\begin{tabular*}{\columnwidth}{@{\extracolsep{\fill}}lrrrrrr@{}}
\hline\hline
& \multicolumn{2}{c}{$2\times32$} & \multicolumn{2}{c}{$3\times56$} & \multicolumn{2}{c}{$4\times84$} \\
Mode & $\epsilon_{\Re}$ & $\epsilon_{\Im}$ & $\epsilon_{\Re}$ & $\epsilon_{\Im}$ & $\epsilon_{\Re}$ & $\epsilon_{\Im}$ \\
\hline
$(2,0,0)$ & 0.001 & 0.007 & $<0.001$ & 0.001 & $<0.001$ & 0.001 \\
$(2,1,0)$ & $<0.001$ & 0.002 & $<0.001$ & 0.001 & $<0.001$ & 0.002 \\
$(2,2,0)$ & 0.005 & 0.997 & 0.001 & 0.238 & $<0.001$ & 0.085 \\
$(3,3,0)$ & 0.001 & 0.340 & $<0.001$ & 0.059 & $<0.001$ & 0.120 \\
$(4,4,0)$ & 0.001 & 0.130 & $<0.001$ & 0.066 & $<0.001$ & 0.025 \\
$(2,2,1)$ & 0.025 & 0.773 & 0.067 & 0.147 & 0.013 & 0.229 \\
$(2,2,2)$ & 0.193 & 0.668 & 0.038 & 0.427 & 0.063 & 0.137 \\
\hline\hline
\end{tabular*}
}
\smallskip
\smallskip
\textit{(b) Architecture timing}\par\smallskip
{\begin{tabular*}{\columnwidth}{@{\extracolsep{\fill}}lrr@{}}
\hline\hline
Architecture & Training step (ms) & Inference (ms) \\
\hline
$2\times32$ & 41.4 & 2.16 \\
$3\times56$ & 79.9 & 3.25 \\
$4\times84$ & 150.8 & 5.97 \\
\hline\hline
\end{tabular*}
}
\end{table}

The size comparison is a numerical robustness test of the trial-function resolution. Within the near-extremal interval shown in Fig.~\ref{fig:architecture_comparison}, the $3\times56$ and $4\times84$ errors remain below the reporting threshold at every spin for every branch. The $2\times32$ errors also remain below it, with the narrow margin noted above. The smaller maximum error of the $4\times84$ representation across the full survey supports its use for the reported spectra, while the timing data quantify the additional numerical cost.

\subsection{Sensitivity to floating-point rounding}
\label{subsec:hardware}

Floating-point addition is not associative, so the same calculation gives slightly different results on different processors or with a different order of parallel reductions~\cite{Goldberg_1991}, and neural-network training inherits these differences even with fixed random seeds~\cite{Pham_2020}. Such differences are harmless for most purposes, but two features of the present problem amplify them: near extremality the residual is almost flat along the damping direction, so a rounding-level change moves the point where the optimizer stops by much more, and each spin starts from the eigenpair accepted at the previous one, so the displacement is carried along the continuation. The frequency solve of Sec.~\ref{subsubsec:frequency_solve} was introduced for this reason: training finds the field and the branch, and the solve then sets the frequency by a stationarity condition (except where step S5 shifts the overtone frequencies).

To measure the remaining sensitivity, we repeated runs with the same random seeds on graphics processing units (GPUs) and on multicore CPUs. Partial repeats agreed within $10^{-3}\%$ at the spins they shared, with one exception of $0.019\%$ (damping of $(2,2,1)$ at $a/M=0$, $4\times84$), and repeated complete $3\times56$ branches of $(4,4,0)$ and $(2,2,1)$ differed by at most $0.093\%$ and $0.034\%$. The largest differences occur for the smallest network at the last spin. We solved $a/M=0.9998$ again from the field accepted at $a/M=0.999$. For $(2,0,0)$, $(2,1,0)$, $(2,2,1)$, and $(3,3,0)$, the $2\times32$ damping agreed within $0.07\%$. For $(2,2,0)$, three processors gave signed damping errors of $+0.997\%$, $+0.93\%$, and $-0.07\%$. This spread, $5\times10^{-5}$ in $M|\Im(\omega)|$, is too small for the residual of this network to resolve. The $(4,4,0)$ and $(2,2,2)$ branches were not repeated.  The $2\times32$ network therefore determines this damping only to about $1\%$; for the two larger networks the differences stay below $0.1\%$.

\subsection{Ringdown observables and waveform consistency}
\label{subsec:ringdown_observables}

The component-wise errors reported above acquire a direct physical interpretation after the dimensionless eigenfrequencies are converted into damped ringdown signals. Following the waveform construction in Ref.~\cite{Luna_2023}, a single real strain component is represented as
\begin{equation}
    \begin{aligned}
        h_{\ell mn}(t)={}&A_{\ell mn}e^{-(t-t_0)/\tau_{\ell mn}}\\
        &\times\cos\!\left[2\pi f_{\ell mn}(t-t_0)+\phi_{\ell mn}\right]\vartheta(t-t_0),
    \end{aligned}
    \label{eq:ringdown_strain}
\end{equation}
\noindent where $A_{\ell mn}$, $\phi_{\ell mn}$, and $t_0$ denote the excitation amplitude, phase, and ringdown start time, respectively, and $\vartheta$ is the unit step function. The eigenvalue problem fixes the oscillation frequency and damping time but not the excitation coefficients. The physical observables, restoring $c$ and $G$, are
\begin{align}
    f_{\ell mn}&=\frac{c^3}{2\pi GM}\Re(M\omega_{\ell mn}),
    &
    \tau_{\ell mn}&=\frac{GM}{c^3|\Im(M\omega_{\ell mn})|},
    \label{eq:physical_frequency_conversion}
\end{align}
\noindent where $M$ is the black-hole mass and $c$ and $G$ are the speed of light and Newton's gravitational constant, respectively. These relations make the mass dependence explicit: $f_{\ell mn}\propto M^{-1}$ and $\tau_{\ell mn}\propto M$. Because the mass enters only as an overall scale, the relative errors quoted in this section are independent of it, so the value $M=200M_{\odot}$ used below is a choice of units rather than a restriction.

\begin{figure*}[!t]
    \centering
    \includegraphics[width=\textwidth]{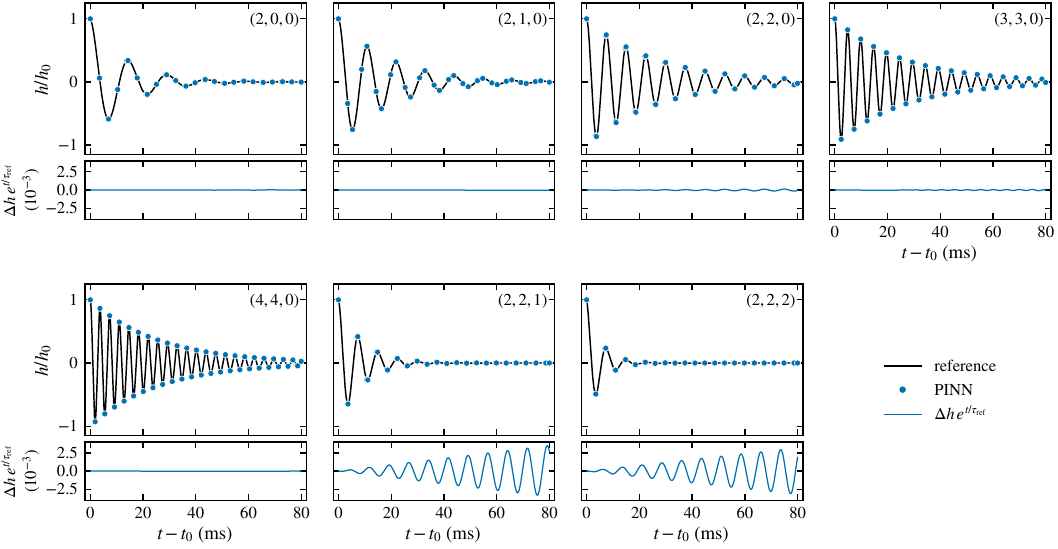}
    \caption{\justifying Single-mode ringdown signals from the PINN frequencies ($4\times84$; markers at turning points) and from the reference frequencies (black curves) at $M=200M_{\odot}$ and $a/M=0.98$, with unit amplitude and zero phase, normalized to the initial amplitude $h_0$. The strips give $\Delta h=h_{\mathrm{PINN}}-h_{\mathrm{ref}}$ divided by the reference envelope $e^{-t/\tau_{\mathrm{ref}}}$, in units of $10^{-3}$.}
    \label{fig:ringdown_waveforms}
\end{figure*}

For the waveform comparison, the reference frequencies are evaluated at the spin nodes. The normalized waveforms set $A_{\ell mn}=1$, $\phi_{\ell mn}=0$, and $t_0=0$. In Fig.~\ref{fig:ringdown_waveforms} we display the ringdown waveforms of all seven branches at $M=200M_{\odot}$ and $a/M=0.98$, each compared with the waveform built from the reference frequencies. The curves are visually coincident, while their residuals resolve the accumulated phase and damping differences. The largest residuals belong to the two overtones, which have the largest absolute frequency errors at this spin (in the damping for $(2,2,1)$ and in the real part for $(2,2,2)$). The narrow panels show $\Delta h$ divided by the reference damping envelope $e^{-t/\tau_{\mathrm{ref}}}$. This normalization matters: since both waveforms decay, the unnormalized difference is bounded by that envelope and would fall to zero at late times even if the two frequencies disagreed appreciably. The normalized difference instead grows with time, as the phase offset accumulates. It grows fastest for $(2,2,1)$, reaching about $3.5\times10^{-3}$ of the envelope after $80$~ms, whereas for the fundamental modes it stays below $1.6\times10^{-4}$. The accumulated disagreement is quantified below by the fitting factor of Eq.~\eqref{eq:ringdown_match}.

The waveform discrepancy is quantified by the noise-weighted fitting factor, namely the match maximized here over relative time and phase~\cite{Cutler_Flanagan_1994,Lindblom_2008}. For two signals $h_1$ and $h_2$, the inner product and fitting factor are
\begin{align}
    (h_1|h_2)&=4\Re\!\int_{f_{\mathrm{low}}}^{f_{\mathrm{Ny}}}
    \frac{\widetilde{h}_1(f)\widetilde{h}_2^{*}(f)}{S_n(f)}\,df,
    \label{eq:ringdown_inner_product}\\
    \mathcal{F}&=\max_{\Delta t,\Delta\phi}
    \frac{(h_1|h_2)}{\sqrt{(h_1|h_1)(h_2|h_2)}},
    \label{eq:ringdown_match}
\end{align}
where $S_n(f)$ is the one-sided detector noise power spectral density, $f_{\mathrm{Ny}}$ is the Nyquist frequency, and $0\leq\mathcal{F}\leq1$, with $\mathcal{F}=1$ for identical signals up to the optimized shifts and $1-\mathcal{F}$ defining the mismatch. The minimum signal-to-noise ratio (SNR) at which this mismatch becomes distinguishable can be estimated as~\cite{Lindblom_2008,Luna_2023},~\cite{Baird_2013}
\begin{equation}
    \rho_{\min}=\left[\frac{D}{2(1-\mathcal{F})}\right]^{1/2},
    \qquad D=4,
    \label{eq:ringdown_snr_min}
\end{equation}
where the $(D=4)$ four parameters are the black-hole mass, spin, signal amplitude, and phase. This threshold is used as a reference-based consistency diagnostic rather than as an event-rate forecast: a large $\rho_{\min}$ means that the PINN and reference waveforms would require a very strong signal to be distinguished, whereas a small value identifies a region in which the residual eigenfrequency error becomes observationally relevant.

\begin{figure*}[!t]
    \centering
    \includegraphics[width=\textwidth]{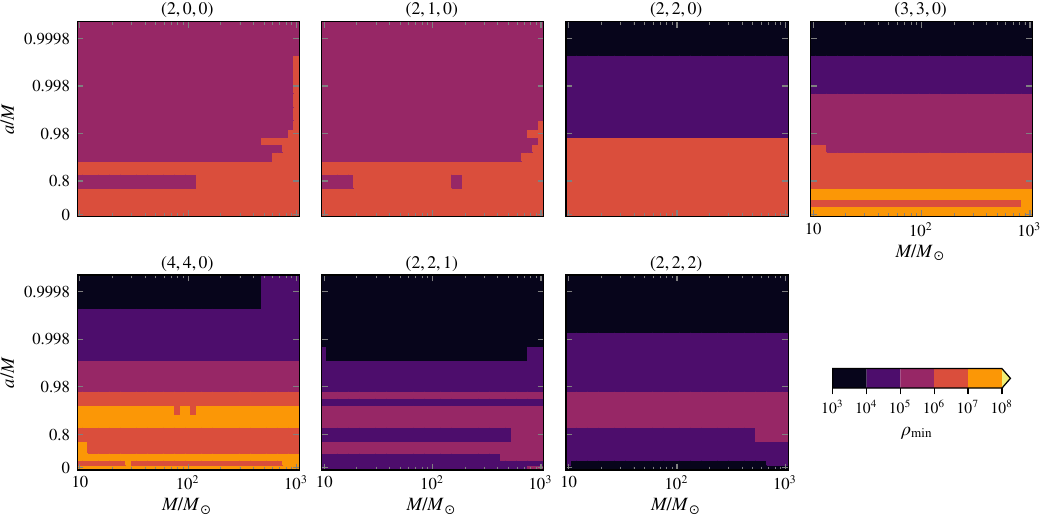}
    \caption{\justifying Minimum SNR $\rho_{\min}$ needed to distinguish the PINN and reference waveforms, binned by decade, over the black-hole mass and the spin (logarithmic in $1-a/M$ and labeled by $a/M$), for the seven branches.}
    \label{fig:ringdown_snr_map}
\end{figure*}

In Fig.~\ref{fig:ringdown_snr_map} we display this threshold for each of the seven branches, over the black-hole mass on the horizontal axis and the spin on the vertical axis, so that every panel shows where in that plane the remaining eigenfrequency error would become observable. The matches are computed with \texttt{PyCBC} version 2.9.0~\cite{PyCBC_2025}, the \texttt{aLIGOZeroDetHighPower} design-sensitivity curve, a one-second window sampled at $16384\,\mathrm{Hz}$, and $f_{\mathrm{low}}=10\,\mathrm{Hz}$. Time and phase shifts are maximized numerically. The sample rate keeps the near-extremal $(4,4,0)$ frequency below the Nyquist limit even at $M=10M_{\odot}$. The map uses all PINN spin nodes through $a/M=0.9998$, without interpolation between trained spins, and spans $10\leq M/M_{\odot}\leq1000$. The horizontal axis is logarithmic in the black-hole mass, with ticks at $M=10$, $100$, and $1000\,M_{\odot}$. To resolve the near-extremal region while preserving the manuscript notation, the vertical plotting coordinate is $-\log_{10}[1-a/M]$ and its tick labels report the actual values of $a/M$; both are dimensionless and hence common to every mass column. The colours classify $\rho_{\min}$ into decade intervals between $10^3$ and $10^8$, rather than implying a continuous colour scale.

The maps in Fig.~\ref{fig:ringdown_snr_map} show that one and the same sub-percent eigenfrequency error has very different waveform consequences from one branch to another: the signal strength needed to tell the two waveforms apart varies by more than two orders of magnitude across the seven branches and across the mass--spin plane. The damped $(2,0,0)$ and $(2,1,0)$ comparisons remain above $\rho_{\min}=2\times10^5$ throughout the displayed domain, and every configuration requires an SNR above $10^3$. Lower thresholds appear primarily in the near-extremal rows, where the long-lived co-rotating modes accumulate phase over many cycles. At the representative point used in Fig.~\ref{fig:ringdown_waveforms}, the seven thresholds range from $1.07\times10^4$ for $(2,2,1)$ to $8.3\times10^5$ for $(2,1,0)$. The minima for $(3,3,0)$ and $(4,4,0)$ are $3.1\times10^3$ and $8.6\times10^3$, respectively; the global minimum, $1.2\times10^3$, occurs for $(2,2,1)$ at $a/M=0.9998$ and $M\simeq14M_{\odot}$. The large thresholds over most of the mass--spin plane provide a direct waveform-level confirmation that the PINN spectrum reproduces the reference solution, while the localized lower-threshold regions identify where tighter eigenvalue accuracy would be required for high-SNR inference.

The comparatively low thresholds of the near-extremal co-rotating modes in Fig.~\ref{fig:ringdown_snr_map} follow from their high quality factors and, for the overtones, from their larger damping errors. They are not a sign that the solver converged to a mode other than the intended $(\ell,m,n)$. Toward extremality, co-rotating Kerr modes become long lived and approach the zero-damping family described in~\cite{Yang_2013}. For two long exponentially damped sinusoids in white noise, with $Q\gg1$ and small relative errors $\delta_R=|\Delta\Re(\omega)|/\Re(\omega)$ in the real frequency and $\delta_I$ in the damping rate, the match maximized over phase is approximately $\mathcal{F}_{\omega}$, with
{\begin{equation}
    \mathcal{F}_{\omega}\simeq\left[1+(Q\delta_R)^2\right]^{-1/2}\left(1-\frac{\delta_I^2}{8}\right),
    \label{eq:high_q_match}
\end{equation}}
\noindent where $Q=\Re(\omega)/\left[2|\Im(\omega)|\right]=\Re(\omega)\tau/2$ is the quality factor of the mode, that is, the number of radians of phase accumulated over one energy damping time, $\tau/2$. The first factor is the loss of phase coherence of two oscillations whose relative phase drifts by about $Q\delta_R$ radians over that time; the second is the overlap $2\sqrt{\gamma_1\gamma_2}/(\gamma_1+\gamma_2)$ of two exponential envelopes whose decay rates $\gamma_{1,2}$ differ by the fraction $\delta_I$, expanded to second order in $\delta_I$. Together they approximate the full noise-weighted fitting factor of Eq.~\eqref{eq:ringdown_match}. With Eq.~\eqref{eq:ringdown_snr_min} this estimate reproduces the lowest thresholds of Fig.~\ref{fig:ringdown_snr_map}. Thus, even a small real-frequency offset is integrated coherently over many cycles when $Q$ is large; the same eigenfrequency accuracy produces a smaller mismatch for shorter-lived modes. Although $Q$ amplifies only $\delta_R$, at the present accuracy both errors matter near extremality: at $a/M=0.9998$ the damping error gives about $84\%$ of the mismatch for $(3,3,0)$ and about half for $(2,2,0)$.

This interpretation also limits the scope of Fig.~\ref{fig:ringdown_snr_map}. The calculation compares isolated, unit-normalized modes with fixed onset and does not predict their excitation amplitudes, correlations in a multimode fit, detector-network response, or source-population distribution. Moreover, Eq.~\eqref{eq:ringdown_snr_min} is an accuracy diagnostic based on the standard small-mismatch interpretation~\cite{Lindblom_2008}; the lowest thresholds therefore indicate where in the mass--spin plane the demand on eigenfrequency accuracy is strongest, and order those regions relative to one another, not as event-rate predictions or parameter-estimation forecasts. Within that scope, the maps show that the PINN frequencies are physically consistent over most of the domain and identify the high-$Q$ corner in which further eigenvalue refinement would matter first.

\begin{figure}[!t]
    \centering
    \includegraphics[width=\columnwidth]{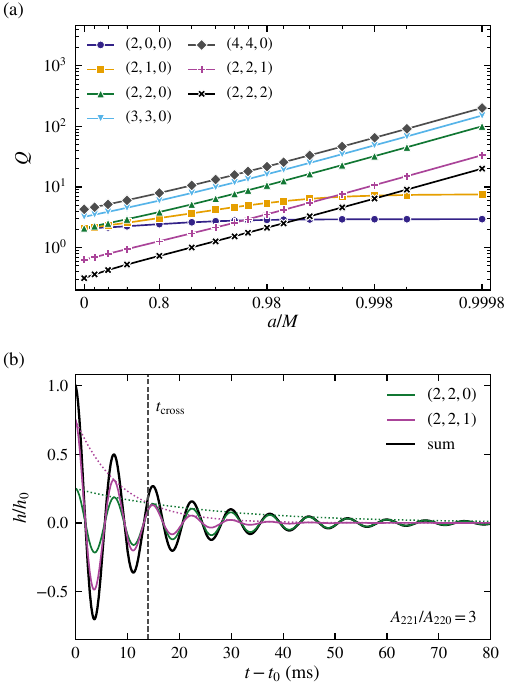}
    \caption{\justifying (a) Quality factors of the seven branches; markers are PINN values ($4\times84$) and curves reference values. (b) Illustrative $(2,2,0)$--$(2,2,1)$ superposition at $M=200M_{\odot}$ and $a/M=0.98$ for $A_{221}/A_{220}=3$, with the component envelopes (dotted) and their crossover (dashed).}
    \label{fig:ringdown_quality_overtone}
\end{figure}

Two additional consequences of the complex frequencies are shown in Fig.~\ref{fig:ringdown_quality_overtone}. Its panel (a) gives the quality factor of each branch as a function of the spin, with the PINN values shown as markers and the reference values as curves; the panel (b) shows the ringdown of a $(2,2,0)$--$(2,2,1)$ superposition at fixed mass and spin, together with the two exponential envelopes and the instant at which they cross. The quality factor is independent of the black-hole mass and is given by
\begin{equation}
    Q_{\ell mn}=\pi f_{\ell mn}\tau_{\ell mn}
    =\frac{\Re(\omega_{\ell mn})}{2|\Im(\omega_{\ell mn})|}.
    \label{eq:ringdown_quality_factor}
\end{equation}

The increasing $Q$ of the co-rotating fundamental branches toward extremality reflects their simultaneous frequency increase and reduced damping. The $(2,2,1)$ branch remains more strongly damped than $(2,2,0)$ and therefore has a smaller quality factor, and $(2,2,2)$ a smaller one still. Nevertheless, an overtone can contribute significantly to the early signal when its excitation amplitude is comparable to that of the fundamental counterpart~\cite{Cheung_2024}. Equating the two amplitude envelopes, $A_{221}e^{-(t-t_0)/\tau_{221}}=A_{220}e^{-(t-t_0)/\tau_{220}}$, gives the crossover time $t_{\mathrm{cross}}$, the instant at which the decaying overtone envelope falls to the level of the fundamental one,
\begin{equation}
    t_{\mathrm{cross}}-t_0=
    \frac{\ln(A_{221}/A_{220})}{\tau_{221}^{-1}-\tau_{220}^{-1}},
    \label{eq:overtone_crossover}
\end{equation}
\noindent provided that $A_{221}>A_{220}$ and $\tau_{221}<\tau_{220}$. Since $t_0$ marks the start of the ringdown, $t_{\mathrm{cross}}-t_0$ measures how long the overtone stays the larger of the two contributions: the signal is dominated by $(2,2,1)$ before that time and by $(2,2,0)$ after it. For the illustrative choice $A_{221}/A_{220}=3$ at $M=200M_{\odot}$ and $a/M=0.98$, the PINN frequencies give $t_{\mathrm{cross}}-t_0\simeq14.0\,\mathrm{ms}$. The amplitude ratio is not predicted by the eigenvalue calculation; it is introduced only to demonstrate the transition from early overtone dominance to late fundamental dominance.

The agreement of the quality factors tests the ratio between the oscillatory and damping parts of each eigenfrequency, rather than either component separately. Together with the waveform matches and the expected fundamental--overtone transition, it shows that the recovered complex frequencies reported in this work preserve the principal physical time scales of Kerr QNM ringdown.

\section{Conclusions}
\label{sec:Conclusions}

Using PINNs, we have solved the gravitational ($s=-2$) Kerr QNM eigenproblem directly on a two-dimensional hyperboloidal domain, without radial--angular separation or an angular separation constant. In order to maintain good accuracy and to be able to identify different modes, we start from the Schwarzschild limit and perform  {a careful} continuation in the Kerr spin. Using the $4\times84$ production network, we have obtained 112 frequencies from seven branches, five fundamental modes and two overtones, from $a/M=0$ through $a/M=0.9998$, with every reference-based real- and imaginary-part error below $0.23\%$ and a median component error of $2\times10^{-4}\%$. Training finds each eigenpair; a Gauss--Newton frequency solve on the polished residual, with the response of the output layer projected out, refines the frequency at every spin, and near extremality the branches with $m\ge2$ are solved in the Cauchy-horizon-fixing gauge. With these, even the two smaller networks stay below $1\%$ error at every spin, and repeating a run on another processor changes the frequencies by less than $0.1\%$ in all but one case.  The recovered decrease in damping toward extremality, the distinction between the $(2,2,0)$ fundamental and the $(2,2,1)$ and $(2,2,2)$ overtones, and the derived quality factors and waveforms show that the computed complex frequencies preserve the principal Kerr ringdown time scales with very good accuracy.

For context, a conventional PINN applied after radial--angular separation reported errors below $1\%$~\cite{Luna_2023}. Hybrid spectral--PINN calculations have reported cumulative real-plus-imaginary relative errors of approximately $0.1\%$ for both soft separated and soft joint formulations, $0.01\%$ for the hard joint formulation, and $0.001\%$ for the hard separated formulation~\cite{Pombo_2026}. The error definitions, mode sets, and spin samples differ, so these values do not define a direct ranking. They nevertheless place the largest present component errors in the same sub-percent regime as the separated conventional PINN and soft joint hybrid calculations, while dedicated spectral eigensolvers and the hard-constrained hybrid formulations reach smaller errors~\cite{Assaad_2025,Pombo_2026}. For observations, the relevant measure is the waveform comparison of Sec.~\ref{subsec:ringdown_observables}: telling a PINN ringdown apart from the reference one requires an SNR above $10^3$ for every branch and spin considered. This accuracy should therefore be adequate for black-hole spectroscopy with next-generation detectors~\cite{Berti_2016}, except perhaps for the loudest signals from remnants close to extremality. 

The significance of this study is methodological rather than a claim of record numerical precision. Apart from the harmonic time and azimuthal reduction and the regularity treatment at the symmetry axis, the neural representation imposes no radial--angular product ansatz, angular separation constant, Leaver-type radial asymptotic prefactor, analytic boundary mask, Chebyshev expansion, or spectral differentiation matrix, which in some cases can be ambiguous \cite{Chung_2024}. The regular field is represented directly in coordinate space, and automatic differentiation supplies the derivatives in the hyperboloidal operator. The principal limitations are the nonlinear optimization cost (Appendix~\ref{app:continuation_details}), the need for continuation and branch diagnostics, the lack of an error estimate independent of reference values beyond the comparison of network sizes, and lower precision than dedicated spectral eigensolvers for the smooth Kerr problem. The horizon-adapted input, the Cauchy-horizon-fixing gauge, the zero-damping predictors, and the hyperboloidal slicing use the horizon structure of Kerr and must be adapted to other backgrounds. The clearest prospective application is a stationary axisymmetric compact-object background supplied as numerical coefficient fields on an irregular or adaptively sampled domain, as may occur for rapidly rotating modified-gravity solutions or compact objects with matter fields~\cite{Luna_2024}. In such settings, a coordinate-space residual method can use scattered or adaptively placed points and interpolated coefficient data without constructing a global expansion or differentiation matrix for the unknown field, and might turn out to be more advantageous over conventional numerical methods. Such applications will require renewed regularity checks, residual certification, and branch diagnostics, particularly for crowded overtone spectra. Nonseparable couplings, such as the gravitational--electromagnetic perturbations of Kerr--Newman black holes~\cite{Dias_2015}, are a natural next test.

\section*{Acknowledgments}

A.F.-S. acknowledges the Valencian Government (CIAICO/2024/111), the Spanish Ministry of Economic Affairs and Digital Transformation (QUANTUM ENIA--Quantum Spain), and the European Union Recovery, Transformation and Resilience Plan--NextGenerationEU (Digital Spain 2026 Agenda). D.D.D. acknowledges the financial support of the Spanish Ministry of Science, Innovation and Universities through the Ram\'on y Cajal programme (grant RYC2023-042559-I), funded by MICIU/AEI/10.13039/501100011033 and by ESF+, of the Spanish State Research Agency (PID2024-159689NB-C21), and an Emmy Noether Research Group funded by the German Research Foundation (DFG) under Grant No. DO 1771/1-1. J.D.M.-G. acknowledges the support listed for A.F.-S. The partial support of KP-06-N62/6 from the Bulgarian science fund is also gratefully acknowledged. R.R.d.A. acknowledges the Spanish Ministry of Science and Innovation (PID2023-148162NB-C22) and the Generalitat Valenciana (PROMETEO/2022/69). Y.V.-G. acknowledges the support listed for A.F.-S. J.A.F. acknowledges the Spanish State Research Agency (PID2024-159689NB-C21; MCIN/AEI/10.13039/501100011033 and ERDF ``A way of making Europe''), the Generalitat Valenciana (CIPROM/2022/49), and the Horizon Europe Staff Exchanges programme (HORIZON-MSCA-2021-SE-01, NewFunFiCO No.~101086251).

\section*{Data availability}
\label{sec:data_and_code}

The training code, the configuration of every mode, the 336 trained networks, and the scripts that reproduce the tables and figures of this work are openly available on Zenodo, doi:10.5281/zenodo.23656788~\cite{Zenodo_2026}.



\appendix
\section{Radial-fixing and Cauchy-horizon-fixing hyperboloidal operators}
\label{app:hyperboloidal_operator}

\counterwithin*{equation}{section}
\renewcommand{\theequation}{\thesection\arabic{equation}}

The starting point is the Teukolsky equation in Boyer--Lindquist coordinates, Eq.~\eqref{eq:intro_teukolsky_operator}, together with the Kerr functions defined there. The hyperboloidal map in Eq.~\eqref{eq:intro_hyperboloidal_map} is completed by the height and azimuthal functions of the radial-fixing minimal gauge~\cite{Assaad_2025},
\begin{align}
    H(\sigma)&{}=-\frac{1}{\sigma}+\frac{2M}{r_+}\ln\sigma
    +\frac{\ln(1-\sigma)}{2r_+\varkappa_+}
    +\frac{\ln(1-\kappa^2\sigma)}{2r_+\varkappa_-},
    \label{eq:app_height}\\
    \bar\chi(\sigma)&{}=\frac{\Omega_+}{2\varkappa_+}\ln(1-\sigma)
    +\frac{\Omega_-}{2\varkappa_-}\ln(1-\kappa^2\sigma),
    \label{eq:app_azimuthal}
\end{align}
\noindent where the surface gravities and angular velocities of the event and Cauchy horizons are
\begin{align}
    \varkappa_+&{}=\frac{1-\kappa^2}{2r_+(1+\kappa^2)},
    &\varkappa_-&{}=-\frac{1-\kappa^2}{2r_+\kappa^2(1+\kappa^2)},
    \nonumber\\
    \Omega_+&{}=\frac{\kappa}{r_+(1+\kappa^2)},
    &\Omega_-&{}=\frac{1}{r_+\kappa(1+\kappa^2)},
    \label{eq:app_horizon_quantities}
\end{align}
\noindent with $M=\tfrac{1}{2}r_+(1+\kappa^2)$. The gauge is called radial-fixing because the radial offset in Eq.~\eqref{eq:intro_hyperboloidal_map} is set to zero, which places the Cauchy horizon at $\sigma_{\rm c}=\kappa^{-2}$ and keeps the constant-$\tau$ surfaces regular there for every subextremal spin~\cite{Assaad_2025}. The coefficient functions of the residual in Eq.~\eqref{eq:pinn_residual} follow from this transformation. We write
\begin{align}
    \mathcal{L}_{1}[h]&=
    \alpha_{2}\partial_{\sigma}^{2}h
    +\alpha_{1}\partial_{\sigma}h
    +\alpha_{0}h
    +\gamma_{2}\partial_{u}^{2}h
    +\gamma_{1}\partial_{u}h,
    \label{eq:app_L1}\\
    \mathcal{L}_{2}[h]&=
    \beta_{1}\partial_{\sigma}h
    +\beta_{0}h.
    \label{eq:app_L2}
\end{align}

We define
\begin{equation}
    \kappa=\frac{a}{r_+},\qquad
    \delta_- = |m-s|,\qquad
    \delta_+ = |m+s|.
    \label{eq:app_notation}
\end{equation}
For the radial-fixing gauge formulated in~\cite{Assaad_2025}, the coefficient multiplying the quadratic eigenvalue term in Eq.~\eqref{eq:pinn_residual} is
\begin{align}
    w={}&4(\sigma+1)
    +\kappa^2\Big[
    -4(1+\kappa^2)^2\sigma^2
    +4\kappa^2\sigma
    +u^2+7
    \nonumber\\
    &\quad
    +4(\kappa^4+2\kappa^2+2)
    +4(\sigma-1)(\kappa^4+\kappa^2+2)
    \Big].
    \label{eq:app_w}
\end{align}
The final contribution inside the square brackets, $4(\sigma-1)(\kappa^4+\kappa^2+2)$, is essential in the Kerr bulk. It vanishes in the Schwarzschild limit and at the event horizon, but omitting it makes the rotating spectrum spuriously Schwarzschild-like even when the residual is small.
The remaining radial and angular coefficients entering Eqs.~\eqref{eq:app_L1}--\eqref{eq:app_L2} are
\begin{align}
    \alpha_2&=\sigma^2(1-\sigma)(1-\kappa^2\sigma),
    \label{eq:app_alpha2}\\
    \alpha_1&=\sigma\left[
    4\kappa^2\sigma^2
    -(1+\kappa^2)\sigma(s+3)
    -2i\kappa m\sigma
    +2(s+1)
    \right],\\
    \alpha_0&=\frac{1}{2}\Big[
    -\delta_-(\delta_++1)
    -\delta_+
    +4\kappa^2\sigma^2
    -2(1+\kappa^2)\sigma(s+1)
    \nonumber\\
    &\quad
    -m^2
    -4i\kappa m\sigma
    +s(s+2)
    \Big],\\
    \beta_1&=
    2\sigma^2
    \left[
    2\kappa^4(\sigma-1)
    +\kappa^2(2\sigma-3)
    -2
    \right]
    +2,\\
    \beta_0&=
    -2\sigma
    \left[
    \kappa^2\{\kappa^2(s+2)+2s+3\}+s+2
    \right]
    \nonumber\\
    &\quad
    +6(\kappa^4+\kappa^2)\sigma^2
    \nonumber\\
    &\quad
    -2im\left[2(1+\kappa^2)\kappa\sigma+\kappa\right]
    +2s(1+\kappa^2-i\kappa u),\\
    \gamma_2&=1-u^2,\\
    \gamma_1&=\delta_- - \delta_+ -u(\delta_-+\delta_+ +2).
    \label{eq:app_gamma1}
\end{align}
The notation used in Eqs.~\eqref{eq:app_alpha2}--\eqref{eq:app_gamma1} is defined in Eq.~\eqref{eq:app_notation}.
The relations in Eqs.~\eqref{eq:app_L1}--\eqref{eq:app_L2}, \eqref{eq:app_notation}, \eqref{eq:app_w}, and \eqref{eq:app_alpha2}--\eqref{eq:app_gamma1} define the operator used in the PINN residual, the relative operator scale in Eq.~\eqref{eq:operator_scale}, and the evaluation snapshots used for post-training diagnostics.

In the Cauchy-horizon-fixing gauge of Sec.~\ref{subsubsec:cauchy_gauge}, used for $m\ge2$ at $a/M\ge0.98$, the operators keep the form of Eqs.~\eqref{eq:app_L1} and \eqref{eq:app_L2} for the field $g(y,u)$, with $y$ in place of $\sigma$, the same $\gamma_2$ and $\gamma_1$, and
\begin{align}
    w&=\frac{4(1+\kappa^2)^2y}{1-\kappa^2}+\kappa^2(u^2+3)+4,
    \label{eq:app_cauchy_w}\\
    \alpha_2&=y^2(1-y),\\
    \alpha_1&=\frac{y}{\kappa^2-1}\Big[\kappa^2\{2(s+1)-y(s+3)\}+2i\kappa my
    \nonumber\\
    &\quad+y(s+3)-2(s+1)\Big],\\
    \alpha_0&=-\frac12\Big[\delta_-(\delta_++1)+\delta_++m^2-\frac{4i\kappa my}{\kappa^2-1}
    \nonumber\\
    &\quad+2y(s+1)-s(s+2)\Big],\\
    \beta_1&=\frac{2}{\kappa^2-1}\Big[-\kappa^4(y-1)^2+\kappa^2(y^2-2y+2)
    \nonumber\\
    &\quad+2y^2-1\Big],\\
    \beta_0&=\frac{2}{\kappa^2-1}\Big[-\kappa^4(y-1)(s+1)+\kappa^2(y-1)
    \nonumber\\
    &\quad+(y-1)s+2y+im\left\{\kappa^3(2y-1)+2\kappa y+\kappa\right\}
    \nonumber\\
    &\quad+is\kappa(1-\kappa^2)u\Big].
    \label{eq:app_cauchy_beta0}
\end{align}
We checked these coefficients against the radial-fixing operator through the exact map of Eq.~\eqref{eq:gauge_map}.

\section{Two-dimensional prefactor method}
\label{app:prefactor_formulation}
\counterwithin*{equation}{section}
\renewcommand{\theequation}{\thesection\arabic{equation}}
Before adopting hyperboloidal coordinates, we implemented a direct two-dimensional formulation in compactified Boyer--Lindquist coordinates. After removing only the time and azimuthal dependence through $\Psi_s=e^{-i\omega t}e^{im\phi}\Phi_s(r,\theta)$, we introduced

\begin{equation}
    x=\frac{r_+}{r}\in(0,1],\qquad u=\cos\theta\in[-1,1],
    \label{eq:app_prefactor_coordinates}
\end{equation}
\noindent so that spatial infinity and the future event horizon correspond to $x=0$ and $x=1$, respectively. The resulting PDE can be written as

\begin{equation}
    C_{xx}\partial_x^2\Phi_s+C_x\partial_x\Phi_s
    +C_{uu}\partial_u^2\Phi_s+C_u\partial_u\Phi_s
    +V\Phi_s=0,
    \label{eq:app_prefactor_original_pde}
\end{equation}
\noindent where the coordinate transformation gives

\begin{align}
    C_{xx}&=\frac{x^4}{r_+^2}C_{rr},
    &C_x&=\frac{2x^3}{r_+^2}C_{rr}-\frac{x^2}{r_+}C_r,
    \nonumber\\
    C_{uu}&=(1-u^2)C_{\theta\theta},
    &C_u&=-uC_{\theta\theta}-\sqrt{1-u^2}\,C_\theta.
    \label{eq:app_prefactor_transformed_coefficients}
\end{align}
The QNM asymptotics and polar regularity are extracted analytically by writing

\begin{equation}
    \begin{aligned}
        &\Phi_s(x,u)=B_x(x)B_u(u)h(x,u),\\
        &p=\partial_x\ln B_x,\qquad q=\partial_u\ln B_u.
    \end{aligned}
    \label{eq:app_prefactor_ansatz}
\end{equation}
Here $B_x$ and $B_u$ are prescribed functions, whereas the complex reduced field $h$ is learned directly on the two-dimensional domain. In particular, Eq.~\eqref{eq:app_prefactor_ansatz} does not impose a factorization of $h$ and does not introduce an angular separation constant. In the open domain, the derivative identities

\begin{align}
    \frac{\partial_x\Phi_s}{B_xB_u}&=\partial_x h+p h,
    \nonumber\\
    \frac{\partial_x^2\Phi_s}{B_xB_u}&=\partial_x^2 h+2p\partial_x h+(p^2+\partial_x p)h,
    \nonumber\\
    \frac{\partial_u\Phi_s}{B_xB_u}&=\partial_u h+q h,
    \nonumber\\
    \frac{\partial_u^2\Phi_s}{B_xB_u}&=\partial_u^2 h+2q\partial_u h+(q^2+\partial_u q)h,
    \label{eq:app_prefactor_derivatives}
\end{align}
\noindent transform Eq.~\eqref{eq:app_prefactor_original_pde} into

\begin{equation}
    H_{xx}\partial_x^2 h+H_x\partial_x h
    +H_{uu}\partial_u^2 h+H_u\partial_u h+V_H h=0,
    \label{eq:app_prefactor_reduced_pde}
\end{equation}
\noindent with

\begin{align}
    H_{xx}&=C_{xx},
    &H_x&=C_x+2pC_{xx},
    \nonumber\\
    H_{uu}&=C_{uu},
    &H_u&=C_u+2qC_{uu},
    \nonumber\\
    V_H&=V+C_{xx}(p^2+\partial_x p)+C_x p
    \nonumber\\
    &\quad+C_{uu}(q^2+\partial_u q)+C_u q.
    \label{eq:app_prefactor_reduced_coefficients}
\end{align}
In the internal $M=1/2$ normalization of the operator, with $r_\pm=M\pm\sqrt{M^2-a^2}$, the explicit factors adapted from the standard Kerr QNM asymptotics~\cite{Leaver_1985,Luna_2023} are

\begin{align}
    B_x(x)&=e^{i\omega r}(r-r_-)^{\rho_-}(r-r_+)^{\rho_+},
    \nonumber\\
    B_u(u)&=e^{a\omega u}(1+u)^{|m-s|/2}(1-u)^{|m+s|/2},
    \label{eq:app_prefactor_explicit_factors}\\
    \rho_-&=-1-s+i\omega+i\zeta_+,
    \nonumber\\
    \rho_+&=-s-i\zeta_+,
    \nonumber\\
    \zeta_+&=\frac{\omega r_+-am}{r_+-r_-}.
    \nonumber
\end{align}
The notation $\rho_\pm$ distinguishes the horizon exponents from the logarithmic derivative $p$. The functions entering Eq.~\eqref{eq:app_prefactor_reduced_coefficients} are therefore

\begin{align}
    p&=\partial_x r\left(i\omega+\frac{\rho_-}{r-r_-}
    +\frac{\rho_+}{r-r_+}\right),
    \nonumber\\
    q&=a\omega+\frac{|m-s|}{2(1+u)}
    -\frac{|m+s|}{2(1-u)},
    \label{eq:app_prefactor_log_derivatives}
\end{align}
\noindent where $r=r_+/x$ and $\partial_x r=-r_+/x^2$. Since $p$ and $q$ are singular at the compactified endpoints, collocation points are kept in the open domain and the preliminary implementation minimized a boundary-vanishing weighted residual with $W=x(1-x)(1-u^2)^2$. The homogeneous amplitude was fixed at one interior point.

This construction successfully produced viable QNM solutions in preliminary calculations and demonstrated that a conventional PINN can address the original two-dimensional equation without separating $h$. The hyperboloidal formulation was nevertheless adopted here because it incorporates the ingoing and outgoing conditions geometrically, keeps the computational field regular at the compactified radial boundaries, and does not require prescribed asymptotic ansätze such as $B_x$ and $B_u$. It therefore provides a more general basis for extensions in which the asymptotic structure is not known in a convenient analytic form.

\section{Numerical details of training and continuation}
\label{app:continuation_details}
\counterwithin*{equation}{section}
\renewcommand{\theequation}{\thesection\arabic{equation}}

This appendix records the numerical choices that support the methodology in Sec.~\ref{sec:Methodology} but are not needed to follow the physical construction. The emphasis is on the role and scale of each setting rather than on a spin-by-spin list of overrides. The numerical settings were calibrated jointly across the proposed modes.

\subsection{Network architecture and resolution}

We compare three network sizes in Table~\ref{tab:architecture_definition}. Unless stated otherwise, the core model  is the $4\times84$ network used for all reported frequency spectra. Sigmoid linear unit (SiLU) activations follow every quadratic layer except for the output layer, whose two channels are returned with no activation. This network has three inputs, two real outputs, and 87,198 neural parameters plus two frequency components. The total is 87,200 adjustable parameters. Each quadratic residual layer is implemented as~\cite{qres_layers},
\begin{equation}
    \mathcal{Q}(z)=Wz+b+(W_{L}z+b_{L})\odot(W_{R}z+b_{R}),
    \label{eq:qres_layer}
\end{equation}
\noindent where $\odot$ denotes component-wise multiplication. The linear branch is Xavier initialized, whereas the quadratic branches start from small weights. The model is therefore initially close to a linear multilayer network and acquires additional nonlinear expressivity during training.

\begin{table}[!t]
\centering
\small
\caption{Quadratic PINN architectures. The configured hidden count is followed by one input-to-width quadratic block; hence the number of activated quadratic blocks is one larger. The total includes the two trainable frequency components. All three networks receive the inputs $(\sigma,u,\eta)$.}
\label{tab:architecture_definition}
\resizebox{\columnwidth}{!}{%
\begin{tabular}{lrrrr}
\toprule
Architecture & Hidden count & Width & Field parameters & Total \\
\midrule
$2\times32$ & 2 & 32 & 6,918 & 6,920 \\
$3\times56$ & 3 & 56 & 29,742 & 29,744 \\
\textbf{$4\times84$} & 4 & 84 & 87,198 & 87,200 \\
\bottomrule
\end{tabular}%
}
\end{table}

\subsection{Baseline setup}

The production model is the $4\times84$ quadratic network defined in Table~\ref{tab:architecture_definition}. Its linear branches are Xavier initialized, while the quadratic branches start close to zero. The collocation cloud includes the compactified boundaries and is resampled every epoch. The remaining settings are grouped by numerical purpose in Table~\ref{tab:app_common_settings}.

\begin{table}[!t]
\centering
\small
\setlength{\tabcolsep}{4pt}
\renewcommand{\arraystretch}{1.10}
\caption{Baseline numerical settings for the production calculations.}
\label{tab:app_common_settings}
\begin{tabularx}{\columnwidth}{@{}p{0.31\columnwidth}X@{}}
\hline\hline
Component & Setting \\
\hline
Optimization & SOAP (Shampoo with Adam in the eigenbasis of the preconditioner)~\cite{SOAP_2025}, field learning rate $10^{-4}$ (reduced to $8\times10^{-5}$ from $a/M=0.9$ and to $6\times10^{-5}$ from $a/M=0.98$), gradient-norm limit $1$ \\
Collocation & $64\times64$ points for $\ell=2$ and $32\times32$ for $\ell=3,4$; exact endpoints included; $30\%$--$52\%$ of the radial points uniform in $\eta$ \\
Residual balance & $w_{\mathrm{rel}}=10^{-2}$; weak mode weight from $5\times10^{-5}$ to $10^{-4}$; residual-gradient weight $10^{-2}$ for $m\ne0$, $a/M\ge0.96$ \\
Training length & $12000$ epochs at $a/M=0$ ($15000$ for the overtones, $30000$ for $(4,4,0)$); $1500$--$12000$ during continuation \\
Learning-rate decay & multiplicative factor $0.5$ during the later training stages \\
Continuation stability & field-only adaptation of $400$--$6200$ epochs followed by frequency and loss-plateau checks \\
Frequency solve (S3) & Gauss--Newton steps of Eq.~\eqref{eq:varpro_step}, at most $40$, relative tolerance $10^{-4}$, step cap $5\%$; Levenberg--Marquardt polish with at most $30$ steps on a $65\times65$ mesh in double precision \\
Overtone stage (S5) & for $n\ge1$ at $0.2\le a/M\le0.96$: $80$ L-BFGS steps of up to $10$ iterations on a $64\times64$ mesh in double precision, kept if the objective decreases on an offset $72\times72$ mesh \\
Final field polish (S6) & at most $30$ Levenberg--Marquardt steps on a $33\times33$ mesh ($64\times64$ for $n\ge1$), $60$ from $a/M=0.97$ for $m\ge2$, $n=0$ and from $a/M=0.8$ for $n\ge1$; kept if the residual decreases on an offset mesh \\
\hline\hline
\end{tabularx}
\end{table}

Near extremality, the radial sampling is shifted toward the horizon and the field learning rate is reduced. These changes resolve the increasingly localized radial structure without modifying the differential operator or introducing an additional boundary condition.

\subsection{Residual scaling and collocation}

The definition in Eq.~\eqref{eq:relative_loss} rescales the local residual by the magnitude of the operator terms. Here $\mathcal{S}_{\Theta}$ is the detached pointwise operator scale,
\begin{align}
    \mathcal{S}_{\Theta}&=|\alpha_{2}\partial_{\sigma}^{2}h|+|\alpha_{1}\partial_{\sigma}h|+|\alpha_{0}h|+|\gamma_{2}\partial_{u}^{2}h|+|\gamma_{1}\partial_{u}h| \nonumber\\&\quad+|\lambda\beta_{1}\partial_{\sigma}h|+|\lambda\beta_{0}h|+|w\lambda^{2}h|.
    \label{eq:operator_scale}
\end{align}

This factor is detached from the computational graph, so $\mathcal{L}_{\mathrm{rel}}$ measures residual imbalance without encouraging the network to reduce the denominator. For a pointwise quantity $f$, the notation $\langle f\rangle_E$ denotes $\sum_p E_p f_p/\sum_p E_p$, with
\begin{equation}
    E(\sigma,u)=\epsilon_{E}+\sigma(1-\sigma)+(1-u^{2}),
    \label{eq:relative_edge_weight}
\end{equation}
\noindent where $\epsilon_E$ and $\epsilon_{\mathrm{rel}}$ are small positive floors. The weight prevents exact endpoints with a small operator scale from dominating the relative residual while retaining them in the collocation set. Thus, $\mathcal{L}_{\mathrm{rel}}$ prevents a small absolute residual from hiding locally large relative errors.

The collocation cloud is regenerated during training rather than kept fixed. At each resampling step, we draw $N_{\sigma}$ radial samples and $N_{u}$ angular samples, and then form the tensor-product cloud

\begin{equation}
    \Omega^{(e)}_{\mathrm{col}}=\left\{(\sigma_i^{(e)},u_j^{(e)})\right\}_{i=1,j=1}^{N_{\sigma},N_u}.
    \label{eq:collocation_cloud}
\end{equation}
The sampling strategy for the set in Eq.~\eqref{eq:collocation_cloud}, including the endpoint bias and the horizon refinement toward extremality, is the one described in Sec.~\ref{subsec:PI_res}.

\subsection{Optimization and trial eigenpairs}

The $\ell=2$ fundamental branches start near the Schwarzschild spectrum, whereas the $(3,3,0)$ and $(4,4,0)$ branches use starting frequencies fixed in advance from the eikonal scaling. For the overtones, the known Schwarzschild frequency, quoted to four digits and held fixed while the field is trained for $15{,}000$ epochs, initializes only the first field optimization. After $a/M=0$, every start is generated from the previously accepted PINN branch, in the near-extremal tail also from predictors built on the analytic zero-damping asymptote, and every candidate is selected by internal residual and branch diagnostics.

The model is trained with SOAP~\cite{SOAP_2025}, using gradient clipping and separate optimizer groups for the field parameters and the two components of $\omega$. Each continuation step begins with the short field-only adaptation stage described above; the frequency is then released and all variables are optimized jointly. Learning-rate decay and early stopping are governed by the stability of the frequency and by the recent loss plateau. Longer adaptation and training windows are used only in the difficult high-spin regimes.

For each spin value, we may train several candidate initializations. The inherited or guided candidate starts from the frequency and weights obtained at the previous spin value. Additional candidates are generated from local secants, short-window history fits, and predictor ensembles formed exclusively from previously accepted PINN frequencies. Predictor sources are assessed by rolling validation on earlier accepted points before the current candidate is trained. In the near-extremal tail some candidates also start from predictors built on the asymptote of Eq.~\eqref{eq:zero_damping_scaling}, anchored on the accepted frequencies, or from a horizon-based predictor (from $a/M=0.94$ for the co-rotating fundamental modes and from $a/M=0.97$ for the overtones). From $a/M=0.6$ the overtone predictors extrapolate $\xi=(\Re\,\omega-m\Omega_{\rm H})/\kappa_{\rm H}$ and $\zeta=\ln(|\Im\,\omega|/\kappa_{\rm H})$, with $\omega=m\Omega_{\rm H}+\kappa_{\rm H}(\xi-ie^{\zeta})$; these variables follow the zero-damping scaling of Eq.~\eqref{eq:zero_damping_scaling} and vary slowly where $\omega$ bends sharply.

\subsection{History-based prediction and field adaptation}

In continuation regimes where the frequency can leave the local branch immediately after being unfrozen, we also use the candidate-specific trust term
\begin{equation}
    \mathcal{L}_{\omega}^{(j)}
    =q_R\left(\frac{\Re(\omega)-\Re(\widehat{\omega}_{k}^{(j)})}{s_{R,k}^{(j)}}\right)^2
    +q_I\left(\frac{\Im(\omega)-\Im(\widehat{\omega}_{k}^{(j)})}{s_{I,k}^{(j)}}\right)^2,
    \label{eq:omega_trust_loss}
\end{equation}
\noindent where $k$ labels the position on the spin grid and $j$ labels one of the candidates trained at the same spin $a_k$; neither index labels a QNM branch, since $(\ell,m,n)$ is fixed throughout each sweep. The quantity $\widehat{\omega}_{k}^{(j)}$ is the candidate-specific frequency proposal computed before training from the previously accepted PINN points. Depending on the candidate, it can come from a local secant, a short history fit, or a predictor ensemble. It supplies the initial value of $\omega$ and, when the trust term is active, its center. The scales $s_{R,k}^{(j)}$ and $s_{I,k}^{(j)}$ quantify the corresponding predictor uncertainty, while $q_R$ and $q_I$ balance the two components. Thus, $\mathcal{L}_{\omega}^{(j)}$ measures how far the trainable frequency moves from the proposal used to launch that particular candidate. The term is justified by the smooth dependence of a QNM branch on the spin: within a single continuation step the physical eigenvalue cannot move far, so a large excursion indicates branch loss rather than improved accuracy.

The coefficients of Eq.~\eqref{eq:omega_trust_loss} are scaled against the remaining physics objective rather than assigned an absolute magnitude. Each continuation step begins with a field-only adaptation stage, after which the frequency is released and the trust contribution is activated. This construction guides local branch tracking without overwhelming the Teukolsky residual.

Let $X_k$ denote either component of the accepted frequency at spin $a_k$. A local continuation estimate has the form

\begin{align}
    d_{X,k-1}&=\frac{X_{k-1}-X_{k-2}}{a_{k-1}-a_{k-2}},
    \nonumber\\
    \widehat{X}_k&=X_{k-1}+(a_k-a_{k-1})\widetilde{d}_{X,k-1},
    \label{eq:app_secant_predictor}
\end{align}
where $\widetilde{d}_{X,k-1}$ is a robustly clipped recent slope. This secant estimate is compared with short linear and quadratic fits to the accepted PINN history, including fits in coordinates adapted to the distance from extremality. Predictor performance is assessed retrospectively on earlier accepted points, and the real and imaginary components may select different sources. At fixed $a_k$, their candidate-specific combination is denoted by $\widehat{\omega}_k^{(j)}$, and their spread supplies the scales $(s_{R,k}^{(j)},s_{I,k}^{(j)})$ in Eq.~\eqref{eq:omega_trust_loss}. The separate symbol $\omega_{h,k}$ denotes the common longer-window quadratic estimate used in the selector. A candidate initialized by that fit can satisfy $\widehat{\omega}_k^{(j)}=\omega_{h,k}$, but the former remains its training proposal and the latter remains the shared post-training reference.

After transfer to a new spin, the frequency is frozen while the field adapts to the new operator. The adaptation interval grows as $a$ approaches $M$, ranging from $400$ to $6200$ epochs in the runs. Once the frequency is released, the trust coefficient is normalized against the current physics objective,

\begin{equation}
    \lambda_{\omega}(e)=g_{\omega}(e)\:
    \frac{\rho\,\operatorname{sg}\!\left(\mathcal{L}_{\mathrm{phys}}\right)}
    {\max\!\left[\operatorname{sg}\!\left(\mathcal{L}_{\omega}\right),10^{-6}\right]},
    \label{eq:app_omega_trust_weight}
\end{equation}
where $g_\omega(e)$ activates the term after the field-only stage, $\operatorname{sg}$ denotes stop-gradient, and the standard guided value is $\rho=1$; at the last spin of $(2,2,0)$ the weight is fixed instead. The normalization keeps the trust term commensurate with the residual objective. Training is stopped only after both frequency components are stable over a trailing window of $300$ epochs and the recent loss has reached a plateau.

\subsection{Candidate selection and overtone control}

At moderate spins, the previous solution provides a suitable starting point for the next value of $a$. Delicate regimes also use nearby candidates generated from the accepted PINN history. At fixed $(\ell,m,n)$ and $a_k$, candidate $j\in\{1,\ldots,N_k\}$ produces a trained pair $(\omega_k^{(j)},h_k^{(j)})$. We define the quantity $\mathcal{S}_k^{(j)}$ as a post-training ranking score used only to decide which of these candidates is selected to start the frequency solve as,

\begin{align}
    \mathcal{S}_{k}^{(j)}&=c_{\mathcal{L}}\mathcal{L}_{k}^{(j)}+c_{\Delta}\mathcal{D}_{\Delta,k}^{(j)}+c_{\mathrm{ov}}\left(1-\mathcal{O}_{k,k-1}^{(j)}\right)\nonumber\\
    &\quad+c_{H}\mathcal{R}_{H,k}^{(j)}\nonumber\\
    &\quad+c_{p}\mathcal{D}_{p,k}^{(j)}+c_{h}\mathcal{D}_{h,k}^{(j)}+c_{R}\mathcal{T}_{R,k}^{(j)}\nonumber\\
    &\quad+c_{I}\mathcal{T}_{I,k}^{(j)}+c_{\max}\mathcal{T}_{\max,k}^{(j)}\nonumber\\
    &\quad+c_{\mathrm{br}}\mathcal{L}_{\mathrm{overtone\;br},k}^{(j)}+c_{\mathrm{sh}}\mathcal{L}_{\mathrm{overtone\;sh},k}^{(j)}\,.
    \label{eq:continuation_score}
\end{align}
This is not a global measure of how good the QNM branch is. Its spin index $k$ identifies the current continuation step, whereas $j$ identifies a candidate at that step. Equation~\eqref{eq:continuation_score} is the expanded form of the grouped score in Eq.~\eqref{eq:continuation_score_overview}.

The terms in Eq.~\eqref{eq:continuation_score} are considered in the order in which they appear. First, $\mathcal{L}_{k}^{(j)}$ is the final training loss of candidate $j$, and

\begin{equation}
    \mathcal{D}_{\Delta,k}^{(j)}=\left|\omega_k^{(j)}-\omega_{k-1}\right|^2
    \label{eq:continuation_step_distance}
\end{equation}
\noindent measures its frequency step from the eigenpair accepted at the preceding spin. The next term, $1-\mathcal{O}_{k,k-1}^{(j)}$, measures the change in the learned field. On a common diagnostic grid, the phase-invariant overlap is

\begin{equation}
    \mathcal{O}_{k,k-1}^{(j)}=\left|\left\langle\frac{h_{k-1}}{\sqrt{\langle |h_{k-1}|^{2}\rangle+\epsilon}},\frac{h_k^{(j)}}{\sqrt{\langle |h_k^{(j)}|^{2}\rangle+\epsilon}}\right\rangle\right|^{2},
    \label{eq:field_overlap}
\end{equation}
\noindent where $\langle\cdot\rangle$ denotes the empirical average. Values near unity indicate a continuous field branch, while smaller values identify a discontinuous eigenfunction change.

The horizon term in Eq.~\eqref{eq:continuation_score}, $\mathcal{R}_{H,k}^{(j)}$, is the mean magnitude of the unweighted Teukolsky residual on the diagnostic subset $\sigma\geq0.92$, with $\sigma=1$ at the horizon. It distinguishes candidates with comparable global losses but different near-horizon residual quality and is not an additional boundary condition.

The predictor-distance term,

\begin{equation}
    \mathcal{D}_{p,k}^{(j)}=\left|\omega_k^{(j)}-\widehat{\omega}_{k}^{(j)}\right|^2,
    \label{eq:predictor_distance}
\end{equation}
\noindent compares the trained frequency with the candidate-specific proposal that initialized it. By contrast, $\omega_{h,k}$ is a single longer-history estimate shared by all candidates at step $k$. It is obtained from a quadratic fit to a wider window of accepted same-branch PINN frequencies and is used only as an independent post-training continuation diagnostic. The two quantities can coincide when a candidate is initialized by that quadratic fit, but their roles remain different: $\widehat{\omega}_{k}^{(j)}$ defines a particular candidate, whereas $\omega_{h,k}$ provides a common reference against which all candidates are compared. With the fit uncertainties $s^{h}_{R,k}$ and $s^{h}_{I,k}$, this comparison is

\begin{equation}
    \mathcal{D}_{h,k}^{(j)}=
    \left(\frac{\Re(\omega_k^{(j)})-\Re(\omega_{h,k})}{s^{h}_{R,k}}\right)^2
    +\left(\frac{\Im(\omega_k^{(j)})-\Im(\omega_{h,k})}{s^{h}_{I,k}}\right)^2.
    \label{eq:history_distance}
\end{equation}
The following two terms quantify reversals of the locally accepted component trends. Defining $\Delta a_k=a_k-a_{k-1}>0$ and

\begin{align}
    \delta_{R,k}^{(j)}&=\operatorname{sgn}(m)\left[\Re(\omega_k^{(j)})-\Re(\omega_{k-1})\right],\nonumber\\
    \delta_{I,k}^{(j)}&=\eta_I\left[\Im(\omega_k^{(j)})-\Im(\omega_{k-1})\right],
    \label{eq:signed_frequency_steps}
\end{align}
\noindent we use the one-sided penalties

\begin{align}
    \mathcal{T}_{R,k}^{(j)}&=\left[\max\!\left(0,\nu_R\Delta a_k-\delta_{R,k}^{(j)}\right)\right]^2,\nonumber\\
    \mathcal{T}_{I,k}^{(j)}&=\left[\max\!\left(0,\nu_I\Delta a_k-\delta_{I,k}^{(j)}\right)\right]^2,
    \label{eq:trend_penalties}
\end{align}
\noindent where $\nu_R$ and $\nu_I$ are nonnegative minimum signed rates per unit spin, so $\nu_R\Delta a_k$ and $\nu_I\Delta a_k$ are the smallest component changes admitted without penalty at that step. For the co-rotating branches, $\eta_I=+1$, so an increasing real part in the co-rotating direction and a less negative imaginary part incur no penalty once their signed steps reach the corresponding local thresholds. A reversal or an insufficient step is penalized quadratically. The next term, $\mathcal{T}_{\max,k}^{(j)}$, places the complementary one-sided cap on an excessively large damping step relative to the range inferred from the accepted history. These trend terms are disabled for $m=0$ and strengthened only in delicate near-extremal regimes. The reason is physical: for the co-rotating branches the real frequency rises and the damping falls monotonically toward extremality, so a reversal indicates a jump to a neighboring branch, whereas the $m=0$ branch has no preferred co-rotating direction. Finally, $\mathcal{L}_{\mathrm{overtone\;br},k}^{(j)}$ checks consistency with the accepted same-overtone history, and $\mathcal{L}_{\mathrm{overtone\;sh},k}^{(j)}$ checks the candidate's radial shape. Their coefficients vanish for fundamental modes.

For nonzero-$m$ branches, selection can be restricted to the low-loss envelope

\begin{equation}
    \mathcal{E}_k(F,N_{\max})=\Big\{j:\operatorname{rank}_{\mathcal{L}}(j)\leq N_{\max},
    \;\; \mathcal{L}_{k}^{(j)}\leq F\mathcal{L}_{k,\min}\Big\}.
    \label{eq:loss_envelope}
\end{equation}
Here $\mathcal{L}_{k,\min}=\min_{1\leq r\leq N_k}\mathcal{L}_{k}^{(r)}$ is the smallest final loss, and $\operatorname{rank}_{\mathcal{L}}(j)=1$ identifies its candidate. The factor $F\geq1$ is the largest admitted ratio $\mathcal{L}_{k}^{(j)}/\mathcal{L}_{k,\min}$, whereas $N_{\max}\leq N_k$ is the largest admitted loss rank; both conditions apply simultaneously. Thus, $F=1.25$ and $N_{\max}=4$ retain at most four candidates with losses within $25\%$ of the minimum. This envelope precedes the composite ranking, restricting $\mathcal{S}_k^{(j)}$ to  candidates with similar convergence so that favorable continuity diagnostics do not compensate for a high final loss. The selected index and pair are given by Eq.~\eqref{eq:candidate_acceptance}. The score is evaluated after independent candidate training and is not backpropagated through the network.

All numerical weights, warm-up lengths, and regime changes were fixed through systematic calibration studies across the retained branches. The calibration sought a common configuration that combined low residuals, stable frequencies, smooth continuation, and continuous fields; mode- or spin-dependent overrides were kept only where these internal diagnostics showed a reproducible loss of robustness; as stated in Sec.~\ref{subsubsec:phys_eqs}, the reference values were also used during this development, to locate the spins at which a version of the method failed. 

The score in Eq.~\eqref{eq:continuation_score} is evaluated after training and is not differentiated through the network. Its nominal scales are summarized in Table~\ref{tab:app_selector_settings}. Near extremality, predictor and trend terms are strengthened within a low-loss candidate envelope (their weights rise to $2\times10^{-2}$ and $0.2$, and the cap $\mathcal{T}_{\max}$ enters with weight $0.5$ to $2$), so branch continuity is considered only among solutions with comparable residual quality. The active settings span $1.25\leq F\leq4$ and $4\leq N_{\max}\leq9$; elsewhere $F=1$ and $N_{\max}=1$, so that only the lowest-loss candidate is eligible, with larger values reserved for difficult high-spin and overtone regimes.

\begin{table}[!t]
\centering
\small
\setlength{\tabcolsep}{4pt}
\renewcommand{\arraystretch}{1.08}
\caption{Grouped nominal weights in the post-training selector.}
\label{tab:app_selector_settings}
\begin{tabularx}{\columnwidth}{@{}>{\raggedright\arraybackslash}X >{\raggedleft\arraybackslash}p{0.22\columnwidth}@{}}
\toprule
Diagnostic & Weight \\
\midrule
Final training loss & $1$ \\
Frequency step from the previous spin & $5\times10^{-2}$ \\
Field overlap and horizon residual & $10^{-3}$ each \\
Agreement with the local predictor & $5\times10^{-3}$ \\
Agreement with the longer history & $10^{-6}$ \\
One-sided frequency-trend penalties & $5\times10^{-2}$ \\
Excess damping step (near-extremal tail) & $0.5$--$2$ \\
\bottomrule
\end{tabularx}
\end{table}

The overtone branches use the same procedure with two weak additions. A radial-shape term discourages collapse to an excessively smooth fundamental-like field, and a same-branch term keeps the candidate close to its own accepted overtone history while the field inherited from the previous spin adapts. The radial-shape weight is of order $10^{-5}$--$10^{-4}$, while the same-branch weight is activated only in the high-spin tail and remains at or below $10^{-3}$. The Schwarzschild overtones receive the longer initialization listed in Table~\ref{tab:app_common_settings}; all later controls depend on the accepted same-overtone history. These numerical controls stabilize optimization and branch propagation without changing the hyperboloidal Teukolsky equation.

\subsection{Convergence of the frequency solve}
\label{app:frequency_solve}

\begin{figure*}[!t]
    \centering
    \includegraphics[width=\textwidth]{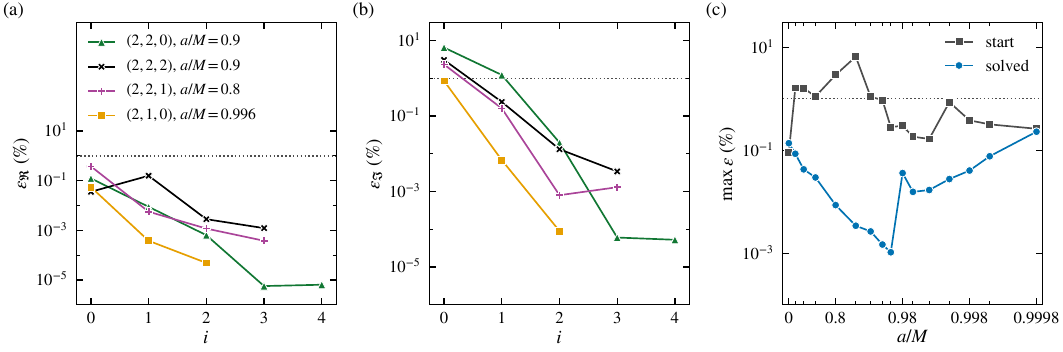}
    \caption{\justifying Convergence of the frequency solve ($4\times84$). (a), (b) Errors $\epsilon_{\Re}$ and $\epsilon_{\Im}$ of Eq.~\eqref{eq:component_error} at outer step $i$, from the starting frequency ($i=0$; extrapolated for $(2,1,0)$ at $a/M\ge0.996$) to the converged one, for the four modes with the largest starting errors, each at its spin of largest starting error. (c) Largest component error over the seven modes at each spin, at the start of the solve and at convergence; the spin axis is logarithmic in $1-a/M$. Step S5 later changes the overtone frequencies at $0.2\le a/M\le0.96$ by at most $0.03\%$. Dotted lines mark $1\%$.}
    \label{fig:frequency_solve}
\end{figure*}

Figure~\ref{fig:frequency_solve} follows step S3 of Algorithm~\ref{alg:pipeline} on the $4\times84$ network, with each outer-step frequency taken from the training logs and its error measured against the reference, which does not enter the iteration. Panels (a) and (b) follow four modes at the spin where their start lies farthest from the reference. The first step removes most of the damping error, from $6.6\%$ to $1.2\%$ for $(2,2,0)$ at $a/M=0.9$, where the $5\%$ cap shortens it, and from $2.3\%$ to $0.16\%$ for $(2,2,1)$ at $a/M=0.8$; after at most three steps every component error of these modes is below $0.004\%$.

Panel (c) gives at each spin the largest of the fourteen component errors: above $1\%$ at six spins before the solve, below $0.23\%$ after it. Where the start already lies close to the reference, the solve can raise an error: at $a/M=0$ the largest error rises from $0.09\%$ (damping of $(2,1,0)$) to $0.14\%$ (damping of $(2,2,2)$, which starts at $0.005\%$), so that the polished residual of this network places the $(2,2,2)$ damping $0.14\%$ from the reference. Near extremality the selected candidates of the modes with $m\ge2$ start from zero-damping predictors anchored on the accepted frequencies (Appendix~\ref{app:continuation_details}; such candidates are selected from $a/M=0.996$ on, although they are available from $a/M=0.94$ or $0.97$); at $a/M=0.9998$ the solve lowers their real-part errors from about $0.02\%$ to at most $0.006\%$ but raises the damping error from below $10^{-4}\%$ to $0.085\%$ for $(2,2,0)$ and from $0.025\%$ to $0.229\%$ for $(2,2,1)$; at this spin panel (c) shows the starting damping error of $(2,1,0)$, which is larger. The near-extremal damping errors of Table~\ref{tab:frequency}(b) are thus set by the polished residual of the network, not by the start of the solve.

The accepted fields do not merely fit the $65\times65$ solve mesh: on the mesh of its cell centers, which is fitted in no step, the mean squared residual of the accepted $4\times84$ fields stays within a factor $0.84$--$1.44$ of its on-mesh value at 110 of the 112 points. The exceptions, $(2,1,0)$ at $a/M=0.999$ and $0.9998$ (factors $14$ and $86$), fit the mesh better than the rest of the domain, yet their frequency errors stay below $0.002\%$.

\subsection{Timing protocol}

The timing entries in panel (b) of Table~\ref{tab:architecture_results} are arithmetic means measured after warm-up in single-precision CPU execution. The timing benchmarks were performed on a MacBook Pro equipped with an Apple M5 chip, a 10-core CPU, and 32~GB of memory. The benchmark keeps the network input and output dimensions, the 4096-point collocation batch, and the automatic-differentiation structure fixed, so these timings characterize architecture-level cost across the modes. Each training measurement includes the residual evaluation, backward pass, and optimizer update, while each inference measurement is a forward pass; 30 training steps and 200 inference passes are averaged. These hardware-dependent measurements quantify the expected increase in computational cost with model size. For scale, a 16-spin branch of $(2,2,2)$ took $4.3$, $5.5$, and $7.4$~h of training on one H100 GPU for the $2\times32$, $3\times56$, and $4\times84$ networks, respectively.

\bibliographystyle{apsrev4-2}
\bibliography{tau}

\end{document}